\documentclass{optica-article}

\journal{opticajournal} 

\articletype{Research Article}

\usepackage{lineno}

\begin{document}

\title{Single-Mode Control of High-Speed and Low-Threshold III-V/Si Quantum Dot Microring Lasers via Azimuthal Gratings}

\author{Xucheng Yang,\authormark{1} Yingtao Hu,\authormark{2} Eunso Shin,\authormark{1} Antoine Descos,\authormark{2} Yatiraj Ramanujam,\authormark{3} Bassem Tossoun,\authormark{2} Yuan Yuan,\authormark{4} Geza Kurczveil,\authormark{2} Jonathan Wierer,\authormark{1}  Ray Beausoleil,\authormark{2} Di Liang,\authormark{3} and Stanley Cheung\authormark{1}}

\address{\authormark{1}North Carolina State University, Department of Electrical and Computer Engineering, 2410 Campus Shore Dr., Raleigh, NC 27606, USA\\
\authormark{2}Hewlett Packard Labs, Large-Scale Integrated Photonics Laboratory, 820 N. McCarthy Blvd., Milpitas, CA 95035, USA\\
\authormark{3}University of Michigan, Ann Arbor, Department of Electrical and Computer Engineering, 1301 Beal Ave., Ann Arbor, MI 48109, USA\\
\authormark{4}Northeastern University, Department of Electrical Engineering, 500 MacArthur Blvd., Oakland, CA 94613, USA
}

\email{\authormark{*}scheung3@ncsu.edu} 


\begin{abstract*} 
Hybrid III–V/silicon quantum-dot microring lasers are compact, energy-efficient O-band sources, but their whispering-gallery cavities are inherently multimode and bidirectional, producing unstable mode hopping that is incompatible with dense wavelength-division multiplexing. We show that an azimuthal grating patterned into the silicon ring — a single lithographic degree of freedom — converts this multimode cavity into a wavelength-addressed, single-mode source. A coupled-mode analysis derives the angular-momentum selection rule from first principles and shows that the inner-wall corrugation replaces the degenerate counter-propagating pair with symmetric and anti-symmetric standing-wave supermodes of unequal radiative loss. At the second-order Bragg condition the anti-symmetric mode is symmetry-protected, yielding a high-quality-factor state at exactly one azimuthal order; finite-element simulations confirm this and identify grating depth as the primary loss-engineering handle. Devices fabricated in-house on a 100 mm silicon-on-insulator platform hold a single longitudinal order with a side-mode suppression ratio of 37.9 dB and continuous, hop-free tuning, while the emission wavelength stays fixed across a factor-of-two change in cavity loading, set lithographically rather than by the gain peak. Because the grating decouples the lasing wavelength from the quantum-dot gain, the detuning becomes a mask-level design variable that sets the temperature of minimum threshold current, reaching 1.95 mA near 50 °C. Combined with side-mode suppression beyond 37 dB and multi-gigahertz direct modulation, these lasers are practical building blocks for cascaded, wavelength-addressed transmitter arrays in data communication and co-packaged optics.

\end{abstract*}

\section{Introduction}
Microring lasers (MRLs) have attracted significant attention as compact, energy-efficient sources for photonic systems, optical logic, and neuromorphic computing, offering advantages over larger conventional linear-cavity lasers such as Fabry–Perot (FP), distributed Bragg reflector (DBR), and distributed feedback (DFB) devices. Nevertheless, the whispering-gallery-mode (WGM) nature of MRL cavities inherently supports bidirectional propagation, which can lead to chaotic multimode lasing behavior and unstable mode hopping \cite{Liang2016_NP,Cheung2025_NC,Zhang2019_Optica,Wan_2018_ACS,Wan_Optica_2017,Wan_PR_2018, Liang_JSTQE_2011,Zhang_JSTQE_2011, Liang_PJ_2011,Cheung_ISLC_2024,Cheung_OFC_2025, Spuesens2011_G4, Sui2015_PR,Campenhout2007_OE,Campenhout2008_PTL}. In this work, we present the implementation of azimuthal gratings within an O-band III-V/Si quantum-dot (QD) MRL cavity to lift the degeneracy between counter-propagating modes while enabling robust single-mode lasing. Achieving this stable operation is critical for the development of compact, low-power DWDM light sources integrated on heterogeneous silicon photonic platforms. A fundamental limitation of conventional microring resonators is their multi-longitudinal-mode nature: the free spectral range (FSR) is finite, and in most practical geometries multiple resonant modes fall within the gain bandwidth of an amplifier or within the bandwidth of interest for a filter. Achieving single-mode operation therefore requires either an extremely small resonator (large FSR), which incurs high bending loss and reduces the quality factor ($Q$-factor), or an additional frequency-selective loss mechanism that suppresses all modes except one \cite{Little1997_JLT,Matsko2006_JSTQE}. Azimuthal diffraction gratings—periodic corrugations inscribed around the circumference of a ring resonator—offer a physically transparent and lithographically compatible approach to mode-selective loss engineering. The grating introduces a spatially periodic perturbation $\Delta \epsilon (r,\phi)$ of the ring's permittivity profile, which, through an angular momentum selection rule, preferentially couples specific whispering-gallery mode (WGM) orders to radiation or to other guided modes \cite{Zhu_OE_2021,Jin_JOSAB_2016,Cai_Science_2012,Chen_ACS_2024}. By designing the grating wavevector to phase-match all modes $\nu \neq \nu_0$ to the radiation continuum while keeping the target mode $\nu_0$ in a momentum gap, one can in principle achieve arbitrarily large mode-selectivity ratios limited only by fabrication disorder and the finite number of grating periods.

\begin{figure}[htbp]
\centering\includegraphics[width=12cm]{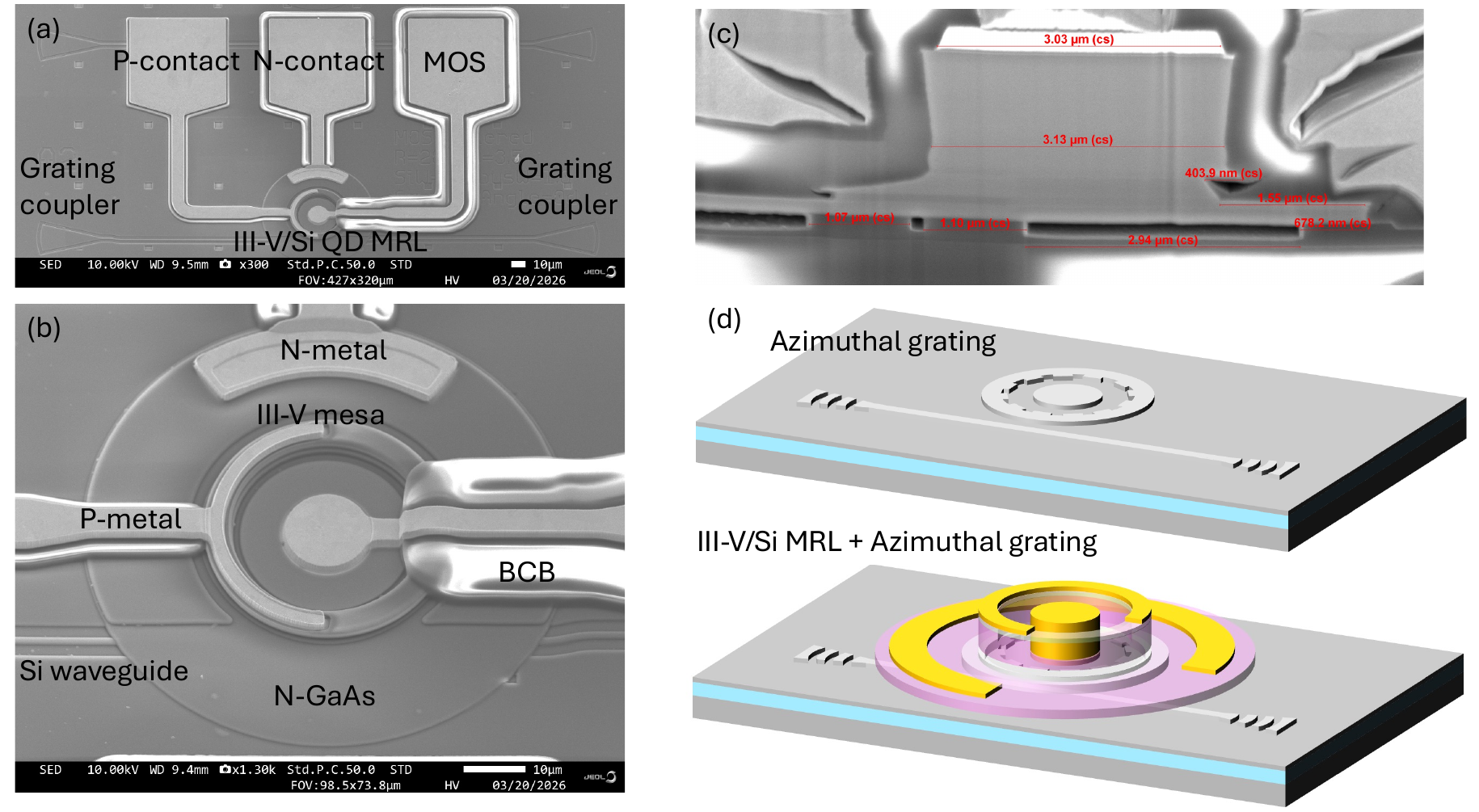}
\caption{(a) Top-view of III-V/Si MRL with azimuthal gratings in silicon. (b) Close up of MRL region. (c) SEM cross-section of ring-coupler region. (d) 3-D schematic of silicon azimuthal gratings with bonded III-V epitaxial stack.}
\label{Fig_SEMandSchematic}
\end{figure}

The fabricated III-V/Si 5QD MRL with an integrated azimuthal grating patterned within the silicon ring resonator is shown in Fig. \ref{Fig_SEMandSchematic}a-d. The azimuthal grating is patterned within the inner circumference of the silicon ring resonator as opposed to the III-V active region in an effort to realize low-threshold current devices by avoiding any recombination sites from sidewall damage. It should also be noted that QDs provide intrinsic protection from these Shockley-Read-Hall (SRH) non-radiative recombination sites by drastically shortening the lateral carrier diffusion due to tightly localized carriers of the three dimensional QD structures \cite{Wan_2018_ACS}. 

\begin{table}[ht]
\centering
\fontsize{7pt}{7pt}\selectfont
\caption{Performance metrics of state-of-the-art MRLs with azimuthal gratings}
\begin{tabular}{l c c c c c c c c c}
\hline
\textbf{Authors} & \textbf{Active region} & \textbf{Platform} & 
\begin{tabular}[c]{@{}c@{}}\textbf{Bus} \\ \textbf{waveguide}\end{tabular} & 
\begin{tabular}[c]{@{}c@{}}\textbf{Wavelength} \\ \textbf{[nm]}\end{tabular} & 
\begin{tabular}[c]{@{}c@{}}\textbf{Radius} \\ \textbf{[$\mu$m]}\end{tabular} & 
\begin{tabular}[c]{@{}c@{}}\textbf{SMSR} \\ \textbf{[dB]}\end{tabular} & 
\begin{tabular}[c]{@{}c@{}}$\mathbf{I_{th}}$ \\ \textbf{[mA]}\end{tabular} & 
\begin{tabular}[c]{@{}c@{}}$\mathbf{P_{out}}$ \\ \textbf{[mW]}\end{tabular} & 
\begin{tabular}[c]{@{}c@{}}$\mathbf{f_{3dB}}$ \\ \textbf{[GHz]}\end{tabular} \\
\hline
S. Zhu \cite{Zhu_OE_2021} & 7QD & III-V/Si & No  & 1300--1370 & 15   & 49 & 15 & 0.075 & NA \\
A. Arbabi \cite{Arbabi_OE_2015} & 1QW & III-V & Yes & 980--990 & 34.5 & 27 & 25.5 & NA & NA \\
L. Feng \cite{Feng_Science_2014} & MQW & III-V & No & 1500 & 4.45 & 14 & $\dag$ & $\ddag$ & NA \\
\textbf{This work} & 5QD & III-V/Si & Yes & 1316--1318 & 25 & 37.9 & 2.25 & 0.48 & 3.5\\
\hline
\end{tabular}
\label{tab:mrl_comparison}

$\dag$ 600 MW/cm$^2$.\\
$\ddag$ $12 \times 10^{4}$ counts.\\
\end{table}

Current state-of-the-art single wavelength MRLs are itemized in Table \ref{tab:mrl_comparison} and benchmarks our device against representative single-mode MRLs employing azimuthal or grating-based mode selection. Zhu et al. \cite{Zhu_OE_2021} achieve the highest side-mode suppression ratio (SMSR = 49 dB) with a 7QD III-V/Si ring, but the device omits bus-waveguide integration and exhibits a comparatively high threshold current of 15 mA with sub-100 $\mu$W output. Arbabi et al. \cite{Arbabi_OE_2015} demonstrate a grating-integrated single-mode MRL with an output bus waveguide, though on an all III-V quantum-well platform operating at 980–990 nm with a 27 dB SMSR and a 25.5 mA threshold. Feng et al. \cite{Feng_Science_2014} realize single-mode operation via parity-time symmetry breaking in a compact (R = 4.45 $\mu$m) III-V MQW ring, but the device is optically pumped (600 MW/cm² threshold) and reaches only 14 dB SMSR without bus-waveguide coupling. In contrast, this work reports, to our knowledge, the first electrically pumped hybrid III-V/Si 5QD MRL to combine bus-waveguide integration with azimuthal-grating mode selection, achieving a milliamp-scale threshold current of 2.25 mA, a 37.9 dB SMSR, and 0.48 mW of on-chip output power at 1316 nm. Furthermore, we demonstrate high-speed direct gain modulation at $f_{3dB}$ = 3.5 GHz and explore high temperature operation from T = 15 - 65 $^{\circ}$C.

We also provide a framework that establishes a self-contained analytical theory that derives the WGM spectrum of the microring cavity with azimuthal grating perturbations. The incorporation of coupled-mode theory derives the angular momentum selection rule and the coupled-amplitude equations from first principles and allows us to analyze the light-cone momentum-gap condition that governs radiative loss for each eigenmode. Furthermore, we examine the loss rate and $Q$-factor which allows us to define a mode-selectivity figure of merit. The analytical designs and demonstrated devices offer a feasible path towards realizing compact, wavelength-addressable O-band laser sources for dense-wavelength-division multiplexing (DWDM) and co-packaged optics.

\section{Theory and Design of Azimuthal MRLs}

\subsection{III-V/Si Hybrid MRL Design for Low-Threshold Operation}
\begin{figure}[htbp]
\centering\includegraphics[width=13cm]{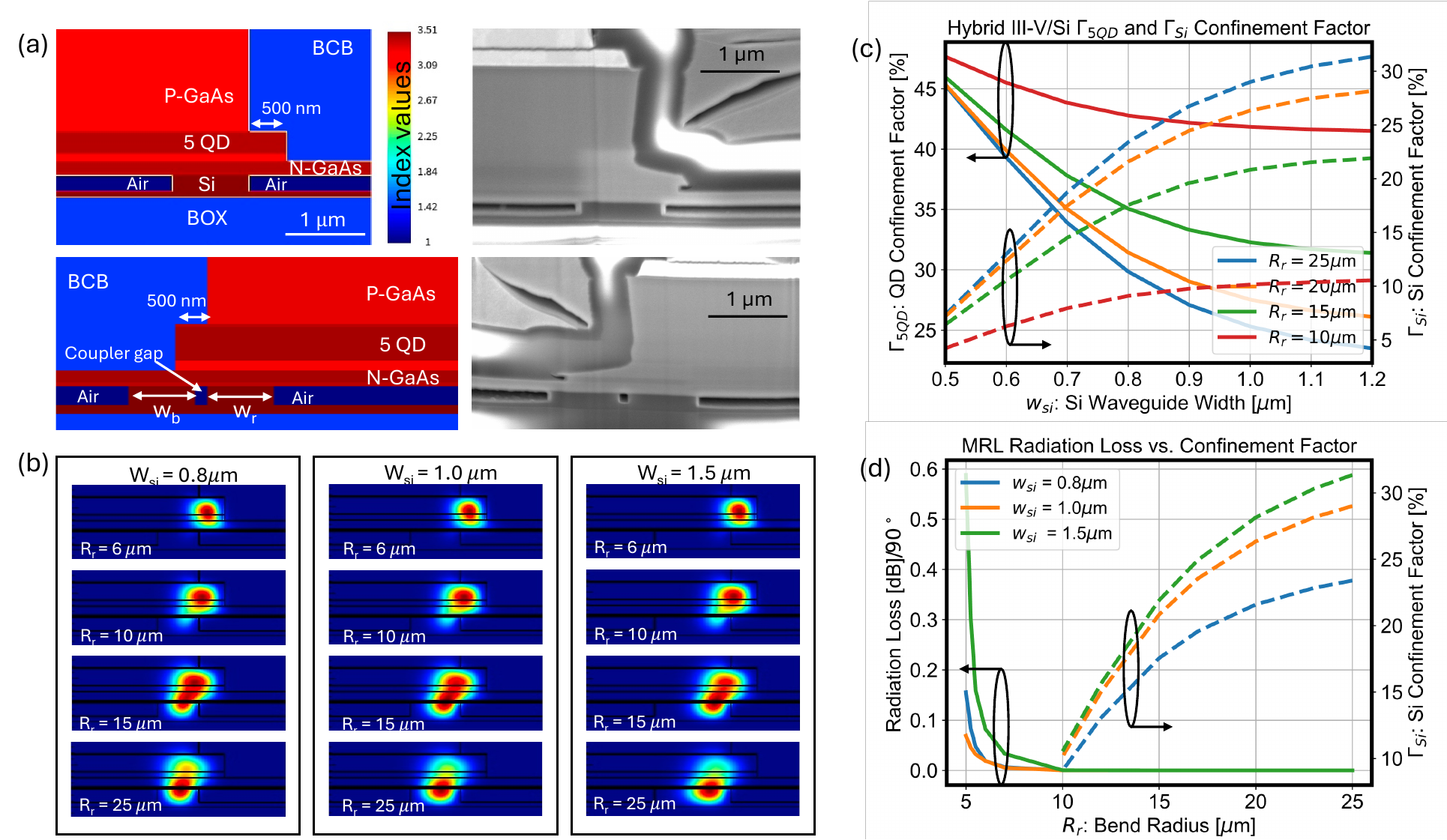}
\caption{(a) Cross-sectional schematics of the III-V/Si MRL mesa and ring--bus coupling
regions with the corresponding refractive indices, together with the matching SEM images. (b) Calculated mode evolution for different silicon waveguide widths $w_{si}$ and bend
radii $R_r$. (c) 5QD and silicon-waveguide confinement factors versus silicon waveguide
width for different ring radii. (d) MRL radiation loss versus silicon confinement factor.}
\label{Fig_ModeSims}
\end{figure}

The hybrid III-V/Si quantum-dot microring laser (MRL) architecture is defined by the combined III-V epitaxial structure and silicon waveguide region, as illustrated in Fig.\ref{Fig_ModeSims}a. As shown in Table \ref{tab:layer_structure}, the epitaxial stack incorporates five layers of InAs/GaAs quantum dots (QDs) with a total active-region thickness of 200 nm and a measured photoluminescence peak near $\lambda = 1289 \pm 5$ nm. The p-type mesa is formed using a 1.54 $\mu$m thick p-AlGaAs cladding layer followed by a heavily doped 100 nm p-GaAs contact layer. On the n-side, the contact region consists of a 150 nm n-GaAs layer together with alternating n-AlGaAs/n-GaAs superlattice layers designed to suppress the propagation of wafer-bonding dislocations. The silicon waveguide was fabricated with a height of 300 nm and an etch depth of 217 nm, where the etch depth was selected primarily to accommodate endpoint-detection tolerances of approximately $\pm 10$ nm. Cross-sectional schematics of both the MRL mesa and coupling regions, including the corresponding refractive indices, are presented in Fig.\ref{Fig_ModeSims}a. To mitigate sidewall scattering and surface carrier recombination, the 5QD active region was laterally offset by 500 nm from the GaAs mesa edge. This offset becomes increasingly important for MRLs employing smaller bend radii, where the optical mode would otherwise exhibit greater overlap with the etched sidewall.

Mode simulations for a ring radius of $R_r = 25 \mu$m (Fig. \ref{Fig_ModeSims}b) further demonstrate that this overlap can be minimized through adjustment of the 5QD confinement factor ($\Gamma_{\mathrm{5QD}}$) by varying the silicon waveguide width. The performance of the hybrid III-V/Si MRL strongly depends on both the confinement factor within the 5QD active region and the coupling coefficient ($\kappa$) between the ring and bus waveguide. Optical confinement calculations obtained using finite-difference eigenmode (FDE) simulations for the 5QD active regions are shown in Fig. \ref{Fig_ModeSims}c and indicate the quantum-dot confinement factor ($\Gamma_{\mathrm{5QD}}$) is highly sensitive to the silicon waveguide width and ring radius $R_r$. Generally, reducing the silicon waveguide width or decreasing the bend radius increases $\Gamma_{\mathrm{5QD}}$ while simultaneously lowering the silicon confinement factor ($\Gamma_{\mathrm{Si}}$). This tradeoff can reduce the laser threshold current ($I_{\mathrm{th}}$), but at the expense of diminished optical output power coupled into the silicon waveguide because of the reduced $\Gamma_{\mathrm{Si}}$. Conversely, excessively small ring radii can degrade the slope efficiency (SE) of the MRL due to enhanced radiation losses. For $R_r < 10~\mu$m, the silicon confinement factor falls below $\Gamma_{\mathrm{Si}} < 10\%$, which can prevent efficient optical coupling into the output silicon waveguide. Consequently, this work primarily investigates devices with ring radii of $R_r = 10$, 15, 20, and 25~$\mu$m together with silicon waveguide widths of $w_{\mathrm{Si}} = 0.8$, 1.0, and 1.5~$\mu$m.

\begin{table}[ht]
\caption{Hybrid III-V/Si 5QD layer structure and material parameters}
\label{tab:layer_structure}
\centering
\fontsize{8pt}{8pt}\selectfont
\begin{tabular}{c l c c c c l}
\hline
\textbf{Layer \#} & \textbf{Layer} & \textbf{Al\%} & \textbf{Thickness [nm]} & \textbf{Doping [cm$^{-3}$]} & \textbf{Index} & \textbf{Purpose} \\
\hline
12 & p-GaAs                 & 0 & 100.0  & $2\times10^{19}$ & 3.406 & Contact \\
11 & p-AlGaAs              & 20 & 20.0   & $2\times10^{19}$ & 3.354 & P-layer\\
10  & p-AlGaAs              & 40 & 1100.0 & $1\times10^{18}$ & 3.245 & P-layer\\
9  & p-AlGaAs              & 40 & 400.0  & $6\times10^{17}$ & 3.245 & P-layer\\
8  & p-AlGaAs              & 20 & 20.0   & $6\times10^{17}$ & 3.354 & P-layer\\
7  & GaAs                  & 0 & 60.0   & n.i.d                & 3.406 & SCH\\
7  & Active Layer (5QD)    & 0 & 200.0  & n.i.d               & 3.406 & Gain \\
7  & GaAs                  & 0 & 20.0   & n.i.d               & 3.406 & SCH\\
6  & n-AlGaAs              & 50 & 100.0  & $6\times10^{17}$ & 3.202 & Etch stop \\
5  & n-GaAs                & 0 & 150.0  & $3\times10^{18}$ & 3.406 & N-layer \\
4  & n-AlGaAs              & 20 & 7.5    & $3\times10^{18}$ & 3.354 & Superlattice \\
3  & n-GaAs                & 0 & 7.5    & $3\times10^{18}$ & 3.406 & Superlattice \\
2  & n-AlGaAs              & 20 & 7.5    & $3\times10^{18}$ & 3.354 & Superlattice \\
1  & n-GaAs                & 0 & 17.5   & $3\times10^{18}$ & 3.406 & Bonding layer \\
\hline
-- & ALD dielectric         & 0 & 22.0   & --                & 1.668 & MOSCAP\\
-- & Si                     & 0 & 300.0  & --                & 3.503 & Waveguide\\
-- & SiO$_2$                & 0 & 2000.0 & --                & 1.447 & Cladding\\
-- & Si substrate           & 0 & --     & --                & 3.503 & -- \\
\hline
\end{tabular}
\end{table}

\noindent In order to optimize low current thresholds $I_{th}$ and reasonable slope-efficiency $SE$ for III-V/Si QD MRL, we need to examine the following equations: 
\begin{equation}
I_{th} = \frac{qV}{\eta_i} \left( BN_{tr}^2 + CN_{tr}^3 e^{\frac{(\alpha_i + \alpha_{rad} + \alpha_m)}{\Gamma_{QD}g_0}} \right) e^{\frac{2(\alpha_i + \alpha_{rad} + \alpha_m)}{\Gamma_{QD}g_0}} , \quad V = \pi t_{QD} \left[ (r_2)^2 - (r_1)^2 \right].
\end{equation}

\begin{equation}
SE = \eta_i \frac{\alpha_m}{\alpha_i + \alpha_{rad} + \alpha_m} \frac{hc}{q\lambda}, \quad \alpha_m = \frac{1}{L_{ring}} \ln \left( \frac{1}{1-k} \right).
\end{equation}

\noindent where $h$, $c$, $q$, $\lambda$, $N_{tr}$, $\Gamma_{QD}$, $g_0$, and $\eta_i$ are the Planck constant, speed of light, unit electric charge, wavelength (1310 nm), transparency carrier density, quantum dot confinement factor, quantum dot material gain, and injection efficiency respectively. The active region volume $V$ is defined by the outer and inner radius of the ring ($r_2$ and $r_1$) respectively where $t_{QD}$ is the total thickness of the 5QD active region (200 nm in this work). $\alpha_i$, $\alpha_{rad}$, $\alpha_m$, $L_{ring}$, and $k$ are the internal active region loss, ring radiation loss, mirror loss, total ring length and power coupling coefficient. $\alpha_i$ in a 5QD InAs/GaAs 
system was reported to be $\alpha_i = 22\text{ cm}^{-1}$ \cite{Amano2006_APL} and will be used in this work. The bimolecular recombination coefficient $B$ and transparent carrier density $N_{tr}$ was taken to be $B = 1.1 \times 10^{-10}\text{ cm}^3/\text{s}$ and 
$N_{tr} = 1.0 \times 10^{18}\text{ cm}^{-3}$ respectively according to work done on 
subthreshold characterization of a 7QD system \cite{Zenari2023_ACS}. The Auger recombination coefficient $C$ is a significant source of nonradiative recombination and have reported values ranging from $C = 4 \times 10^{-29} - 8 \times 10^{-29}\text{ cm}^3/\text{s}$ for 
temperatures from $T = 100 - 300\text{ K}$ \cite{Ghosh2001_APL}. The modal gain for 5QD region was reported to be $\Gamma_{5QD}g_0 = 43\text{ cm}^{-1}$ \cite{Amano2006_APL} for a dot density of $8.0 \times 10^{10}\text{ cm}^{-2}/\text{layer}$ and Innolume reported 
$\Gamma_{7QD}g_0 = 45\text{ cm}^{-1}$ with $\Gamma_{7QD} = 9\%$ for a 7QD region \cite{Maximov2008_SST,Uvin2018_OE}. Based on these reported values, it is reasonable to estimate a material gain 
$g_0 \sim 500\text{ cm}^{-1}$.

\begin{figure}[htbp]
\centering\includegraphics[width=13cm]{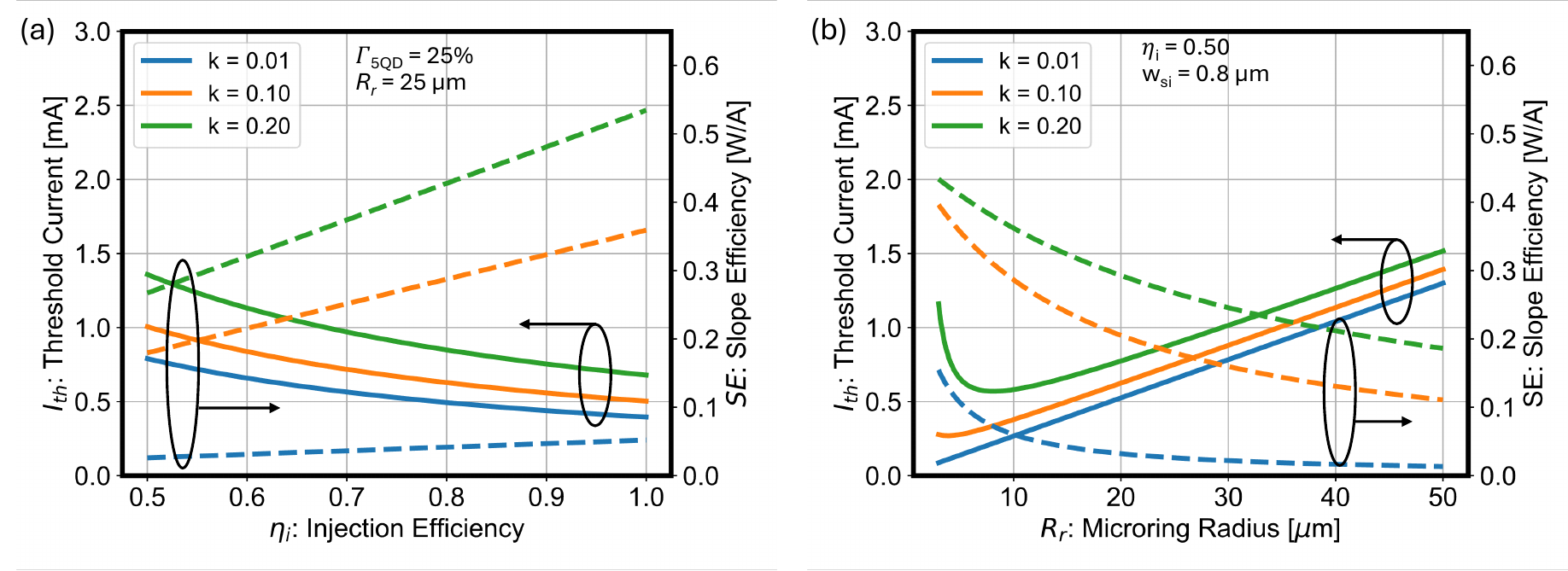}
\caption{Calculated threshold current $I_{th}$ and slope efficiency $SE$ for three different power coupling coefficients $k$ = 0.01, 0.10, 0.20 with (a) a MRL radius of $R_{r}$ = 25 $\mu$m assuming a confinement factor $\Gamma_{5QD} = 25\%$, (b) varying MRL radius $R_{r}$ assuming injection efficiencies of $\eta_i$ = 0.50.}
\label{Fig_MRLdesign}
\end{figure}

As a result, the current threshold $I_{th}$ and slope 
efficiency $SE$ can be determined as a function of injection efficiency $\eta_i$, power 
coupling coefficient $k$, and confinement factor $\Gamma_{5QD}$ as shown in 
Fig.~\ref{Fig_MRLdesign}a. Given a constant MRL radius of $R_r = 25\ \mu\text{m}$ 
and a fixed $\Gamma_{5QD}$, the threshold current $I_{th}$ decreases monotonically 
with increasing $\eta_i$ because of improved carrier injection into the active region. 
A $\eta_i = 0.87$ has been reported in several works \cite{Zenari2023_ACS, Jung2018_ACS}. Sub-mA thresholds of $I_{th} < 1\text{ mA}$ is possible by improving either the modal gain through $\Gamma_{5QD}$ or a higher quality ($Q$) - factor ring via smaller power coupling 
coefficient $k$, albeit at the expense of lower $SE$. In addition, a reduction of 
$R_r$ by $40\%$ can improve the reduction of $I_{th}$ by $33 - 50\%$ for increasing 
values of $\eta_i$. All calculations for Fig.~\ref{Fig_MRLdesign}ab are done 
at $T = 25\ ^\circ\text{C}$ and neglects self-heating but serves as a general guide 
for designing sub-mA threshold III-V/Si QD MRLs.

\subsection{Analytical Theory of Mode-Selective Azimuthal Gratings}

A microring resonator of outer radius $R$, ring width $w$, and effective refractive index $n_{eff}$ supports whispering-gallery modes (WGMs) that are solutions to the scalar wave equation under open-boundary conditions. The resonant wavelength satisfies $\lambda_{\nu} = 2\pi R n_{eff} (\lambda_{\nu})/\nu$. The goal is to engineer the azimuthal grating such that one target mode $\nu$ experiences negligible loss (high-$Q$), while all others are strongly scattered. The transverse-electric (TE) modes are characterized by an integer azimuthal mode number $\nu$ and a radial order $m$, and are written in cylindrical coordinates as:

\begin{equation}
\Psi(r,\phi,t)
=
\sum_{m}
f_{\nu,m}(r)\,e^{-i\omega_m t}
\left[
a_{+,m}(t)e^{i\nu\phi}
+
a_{-,m}(t)e^{-i\nu\phi}
\right].
    \label{eq:wgm}
\end{equation}

\noindent where $f_{\nu}(r)$ is the radial envelope and the $\pm$ signs correspond, respectively, to counter-clockwise (CCW) and clockwise (CW) propagation. For this case, we will assume single-mode transverse propagation such that radial order $m$ = 1. In the absence of any perturbation, the two CW and CCW modes are doubly degenerate:

\begin{equation}
    \omega_{\mathrm{CW}} = \omega_{\mathrm{CCW}} \equiv \bar{\omega}
        = \omega_{r} - i\,\frac{\omega_{r}}{2Q_{0}},
    \label{eq:unperturbed_omega}
\end{equation}

\noindent where $\omega_{r}$ is the real resonance frequency and $Q_{0}$ is the unperturbed quality factor set solely by radiation leakage through the outer wall:

\begin{equation}
    Q_{0} = \frac{\omega_{r}}{2\,|\mathrm{Im}(\omega_{\mathrm{CW}})|}
           = \frac{\omega_{r}}{\gamma_{0}}.
    \label{eq:Q0}
\end{equation}

A general azimuthal grating with $N_g$ periods is modelled as a periodic perturbation of
the ring's permittivity, expanded as a Fourier series:
 
\begin{equation}
    \Delta\varepsilon(r,\phi)
        = \Delta\varepsilon_0(r)
          \sum_{q=-\infty}^{\infty} c_q\, e^{iqN_g\phi},
    \label{eq:grating_perturbation}
\end{equation}
 
\noindent where $\Delta\varepsilon_0(r)$ is the transverse amplitude profile of the
permittivity modulation (non-zero only within the waveguide cross-section), $N_g$ is
the number of grating protrusions around the inner wall, $q = \pm1, \pm2, \ldots$ is
the diffraction order, and $c_q$ are the Fourier coefficients of the grating tooth
profile. For a rectangular tooth profile with duty cycle $\eta$, the coefficients are
$c_q \propto \sin(q\pi\eta)/q$, so that higher diffraction orders are progressively
weaker. For a simple sinusoidal grating, $c_{\pm 1} = 1/2$ and all other $c_q = 0$. We treat $\Delta\varepsilon$ as a perturbation to the unperturbed Helmholtz equation where the perturbed wave equation for the field amplitude is: $\nabla^2 E + k_0^2\left[\varepsilon_0(r) + \Delta\varepsilon(r,\phi)\right]E = 0$. Projecting onto the fundamental radial mode via the orthogonality relation
$\iint f_\nu(r)\,f_{\nu'}(r)\,r\,\mathrm{d}r\,\mathrm{d}z = \delta_{\nu\nu'}$,
the coupling coefficient between azimuthal modes $\nu$ and $\nu'$ is:
\begin{equation}
    \kappa_{\nu\nu'} =
        \frac{\omega_r}{4}
        \iint f^*_{\nu'}(r)\,\Delta\varepsilon_0(r)\,f_\nu(r)
        \,r\,\mathrm{d}r\,\mathrm{d}z
        \times
        \frac{1}{2\pi}\sum_{q=-\infty}^{\infty} c_q
        \int_0^{2\pi}
            e^{i(\nu - \nu' + qN_g)\phi}
        \,\mathrm{d}\phi.
    \label{eq:coupling_coeff}
\end{equation}
 
\noindent The angular integral for each diffraction order $q$ is evaluated in closed
form using the orthogonality of complex exponentials:
 
\begin{equation}
    \frac{1}{2\pi}\int_0^{2\pi} e^{i(\nu - \nu' + qN_g)\phi}\,\mathrm{d}\phi
        = \delta_{\nu',\,\nu + qN_g},
        \qquad q = \pm1,\,\pm2,\,\ldots
    \label{eq:ortho_q}
\end{equation}
 
\noindent Substituting Eq.~\eqref{eq:ortho_q} into Eq.~\eqref{eq:coupling_coeff}
yields the fundamental angular momentum selection rule: A grating of order $N_g$ couples
azimuthal mode $\nu$ to modes with azimuthal order $\nu' = \nu + qN_g$ for integer
$q = \pm1, \pm2, \ldots$ The coupling strength of each order is weighted by the
Fourier coefficient $c_q$ of the grating tooth profile. All other couplings vanish
identically. This reproduces the phase-matching condition $\nu' = \nu - qN_g$ from
Eq.~\eqref{eq:phasematch}, with $q = 1$ giving the dominant radiation channel
$\nu' = \nu - N_g$.
 
The transverse overlap integral reduces to a scalar coupling strength:
 
\begin{equation}
    g_q = \frac{\omega_r}{4}\, c_q
        \iint f^*_{\nu'}(r)\,\Delta\varepsilon_0(r)\,f_\nu(r)
        \,r\,\mathrm{d}r\,\mathrm{d}z,
    \label{eq:scalar_coupling}
\end{equation}
 
\noindent so that the full coupling coefficient for each diffraction order is
$\kappa_{\nu\nu'} = g_q$ for $\nu' = \nu + qN_g$.
Since $|c_q|$ decreases with increasing $|q|$, the dominant coupling is at $q = 1$,
which for $N_g = \nu$ gives $\nu' = 0$ --- a purely radially outgoing wave that
escapes the cavity immediately, producing the large radiation loss seen in the
symmetric mode. Distributing $N_{g}$ identical protrusions uniformly along the inner wall at radius $R_{\mathrm{inner}}$ introduces a periodic dielectric perturbation with angular wavenumber $K_{g} = N_{g}/R_{\mathrm{inner}}$. By the angular phase-matching condition~\cite{Jin_JOSAB_2016}, the grating couples a WGM of azimuthal order $\nu$ into radiation modes whose azimuthal orders are

\begin{equation}
    \nu' = \nu - q\,N_{g}, \qquad q = \pm 1,\, \pm 2,\, \ldots
    \label{eq:phasematch}
\end{equation}

\noindent A radiation mode with $|\nu'|$ smaller than $\nu$ has a larger radial propagation constant
and therefore leaks efficiently into the surrounding air, enhancing radiation loss. The grating-induced
extra decay rate for a mode of order $\nu$ can be estimated via first-order perturbation theory
(Fermi's golden rule):

\begin{equation}
    \gamma_{g} \propto
        \left(\frac{d}{R_{\mathrm{inner}}}\right)^{2}
        \left|f_{\nu}(R_{\mathrm{inner}})\right|^{2}
        \rho(\omega_{r}),
    \label{eq:gamma_g}
\end{equation}

\noindent where $d$ is the grating depth and $\rho(\omega_{r})$ is the local density of radiation
states at the resonance frequency.

In the basis of the CW and CCW travelling-wave amplitudes $a_{+}$ and $a_{-}$, the inner-wall grating
acts simultaneously as (i) an intra-mode radiation scatterer (rate $\gamma_{g}$) and (ii) an
inter-mode back-scatterer (coupling rate $\kappa$). The time-domain coupled-mode equations read:

\begin{equation}
\frac{d a_{+}}{dt}
=\underbrace{-i\omega_r a_{+}}_{\text{oscillation}}-\underbrace{\frac{\gamma_0+\gamma_g}{2}a_{+}}_{\text{loss}}-\underbrace{i\kappa a_{-}}_{\text{backscattering}}
\label{eq:cme_cw}
\end{equation}

\begin{equation}
\frac{d a_{-}}{dt}=\underbrace{-i\omega_r a_{-}}_{\text{oscillation}}-\underbrace{\frac{\gamma_0+\gamma_g}{2}a_{-}}_{\text{loss}}-
\underbrace{i\kappa a_{+}}_{\text{backscattering}}
\label{eq:cme_ccw}
\end{equation}

\begin{equation}
\frac{d}{dt}
\begin{bmatrix}
a_{+} \\
a_{-}
\end{bmatrix}
=
\begin{bmatrix}
-i\omega_r-\dfrac{\gamma_0+\gamma_g}{2} & -i\kappa \\
-i\kappa & -i\omega_r-\dfrac{\gamma_0+\gamma_g}{2}
\end{bmatrix}
\begin{bmatrix}
a_{+} \\
a_{-}
\end{bmatrix}.
\end{equation}

\noindent Both CW and CCW components experience the same radiation losses at this stage because
Eq.~\eqref{eq:phasematch} applies equally to $e^{+i\nu\phi}$ and $e^{-i\nu\phi}$. Equations~\eqref{eq:cme_cw}--\eqref{eq:cme_ccw} are diagonalised by the symmetric and anti-symmetric superpositions:
\begin{align}
    |S\rangle &= \frac{a_{+}+a_{-}}{\sqrt{2}} \propto \cos(\nu\phi),
    \label{eq:sym}\\[6pt]
    |A\rangle &= \frac{a_{+}-a_{-}}{\sqrt{2}} \propto \sin(\nu\phi).
    \label{eq:asym}
\end{align}

\noindent where their complex eigenfrequencies are derived as:
\begin{align}
    \omega_{S} &= \omega_{r} + \kappa
                  - i\,\frac{\gamma_{0}+\gamma_{g}^{S}}{2},
    \label{eq:omega_S}\\[6pt]
    \omega_{A} &= \omega_{r} - \kappa
                  - i\,\frac{\gamma_{0}+\gamma_{g}^{A}}{2},
    \label{eq:omega_A}
\end{align}

\noindent which yield the quality factors $Q_S$ and $Q_A$ for the symmetric and anti-symmetric respectively:

\begin{equation}
    Q_{S} = \frac{\omega_{r}}{\gamma_{0}+\gamma_{g}^{S}},
    \qquad
    Q_{A} = \frac{\omega_{r}}{\gamma_{0}+\gamma_{g}^{A}}.
    \label{eq:QSA}
\end{equation}

The two grating-induced loss rates $\gamma_{g}^{S}$ and $\gamma_{g}^{A}$ are not equal;
their values are governed by the symmetry selection rule derived in the following section.
Because each standing wave contains both counter-propagating travelling-wave components, its
coupling to the $\nu'=0$ radiation channel at the Bragg condition $N_{g}=\nu$ involves
both first-order diffraction terms of Eq.~\eqref{eq:grating_perturbation}: the $e^{+i\nu\phi}$
component reaches $\nu'=0$ via $q=+1$, while the $e^{-i\nu\phi}$ component reaches the same
channel via $q=-1$ [cf.\ Eq.~\eqref{eq:phasematch}]. The coupling of each standing wave into
this shared channel is therefore proportional to the squared modulus of the overlap integral:
\begin{equation}
    I_{S,A} = \int_{0}^{2\pi} E_{S,A}(\phi)\,
    \tfrac{1}{2}\!\left(e^{iN_{g}\phi}+e^{-iN_{g}\phi}\right)d\phi
    = \int_{0}^{2\pi} E_{S,A}(\phi)\,\cos(N_{g}\phi)\,d\phi,
    \label{eq:overlap}
\end{equation}
\noindent where $E_{S}(\phi)\propto\cos(\nu\phi)$ and $E_{A}(\phi)\propto\sin(\nu\phi)$; the
angular origin $\phi=0$ is taken at the center of a grating tooth, so that the two first-order
coefficients of Eq.~\eqref{eq:grating_perturbation} are real and equal ($c_{-1}=c_{+1}$, since
$\Delta\varepsilon$ is real), with their constant magnitude absorbed into the radial overlap
$g$ below. The power
radiated is proportional to $\left|I_{S,A}\right|^2$, therefore the grating-induced decay rate is:
\begin{equation}
    \gamma_{g}^{S,A}=\frac{\omega_r}{4}\,\underbrace{\iint f_{\nu}^{*}(r)\,\Delta\varepsilon_{0}(r)\,f_{\nu}(r)\,r\,dr\,dz}_{=\,g\;(\text{radial overlap})}\,\cdot
    \left| I_{S,A} \right|^{2}.
    \label{eq:decay}
\end{equation}
The section below explores the analysis of Eqs.~\eqref{eq:overlap}--\eqref{eq:decay},
giving insight into single-mode behavior.
 
\paragraph{Symmetric mode.}
Substituting $E_{S}$ and using the product-to-sum identity:
\begin{equation}
    I_{S} = \int_{0}^{2\pi} \cos(\nu\phi)\cos(N_{g}\phi)\,d\phi
           = \frac{1}{2}\int_{0}^{2\pi}
             \!\left[\cos\!\bigl((\nu+N_{g})\phi\bigr) + \cos\!\bigl((\nu-N_{g})\phi\bigr)\right]d\phi.
    \label{eq:IS}
\end{equation}
\noindent When $N_{g} = \nu$, the second cosine has zero argument and the integral evaluates
to $\pi$, giving $I_{S} = \pi \neq 0$. Hence the symmetric mode couples strongly to the radiation
field:
\begin{equation}
    \gamma_{g}^{S} =\frac{\omega_r}{4}\, g \cdot |I_S|^2=\frac{\omega_r}{4}\, g \cdot \pi^2> 0 \qquad (N_{g} = \nu).
    \label{eq:gammaS}
\end{equation}
 
\paragraph{Anti-symmetric mode.}
Substituting $E_{A}$:
\begin{equation}
    I_{A} = \int_{0}^{2\pi} \sin(\nu\phi)\cos(N_{g}\phi)\,d\phi
           = \frac{1}{2}\int_{0}^{2\pi}
             \!\left[\sin\!\bigl((\nu+N_{g})\phi\bigr) + \sin\!\bigl((\nu-N_{g})\phi\bigr)\right]d\phi
           = 0.
    \label{eq:IA}
\end{equation}
\noindent Every sine integrates to zero over a full round trip for any integer argument---including
$N_{g}=\nu$, where $\sin\bigl((\nu-N_{g})\phi\bigr)$ vanishes identically---so $I_{A}=0$ exactly,
for all $\nu$. The anti-symmetric mode is therefore \emph{symmetry-protected} from this
radiation channel:
\begin{equation}
    \gamma_g^{A}=\frac{\omega_r}{4}\, g \cdot |I_A|^2 = 0 \qquad (N_{g} = \nu),
    \label{eq:gammaA}
\end{equation}
\noindent and its quality factor reverts to the unperturbed value, $Q_{A} \approx Q_{0}$.

\begin{figure}[htbp]
\centering\includegraphics[width=11cm]{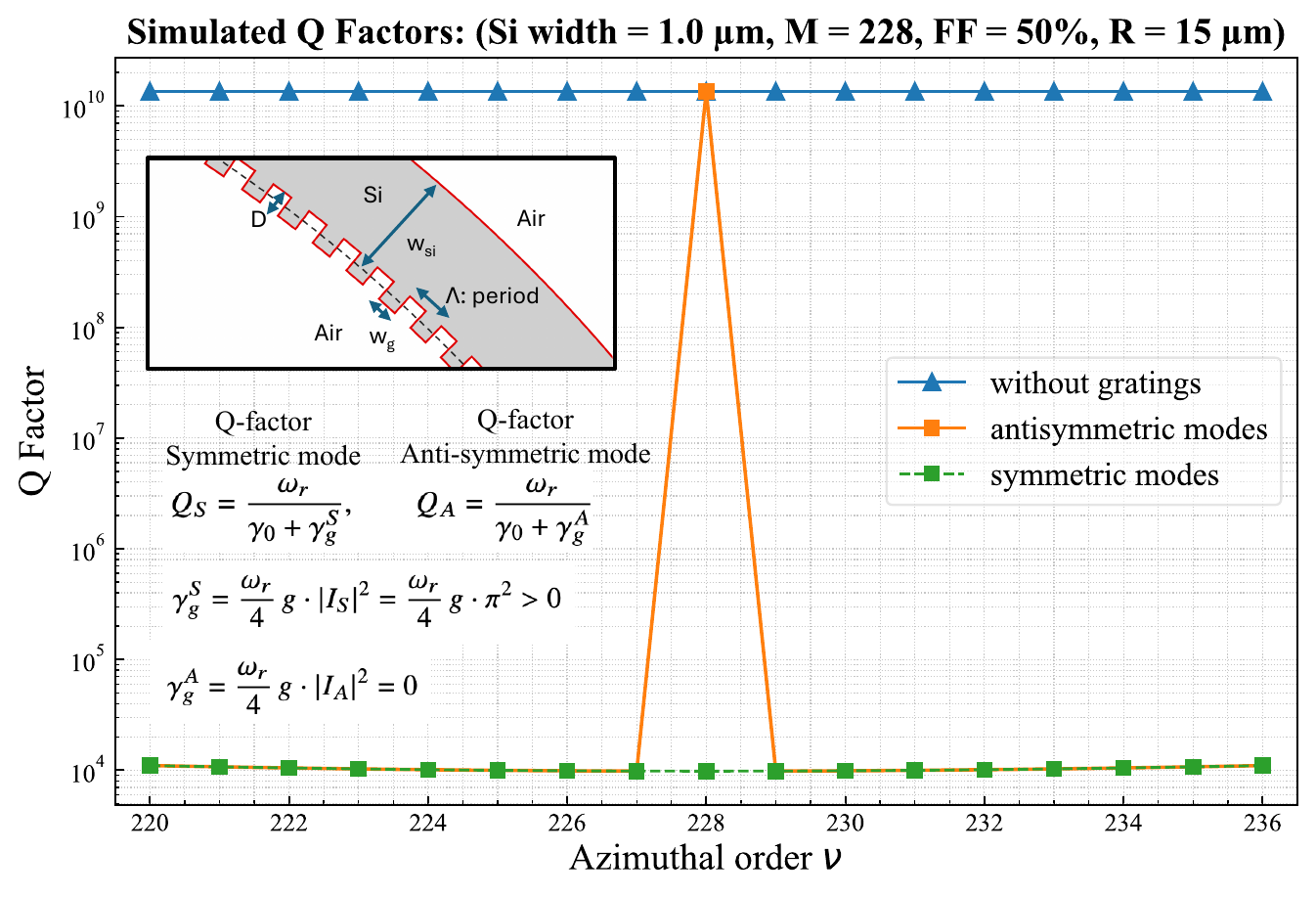}
\caption{Analytical results for $2^{nd}$ order azimuthal gratings where $\nu\lambda/n_{\text{eff}} = 2\pi R$}
\label{Fig_AzimuthalModelAnalytical}
\end{figure}

Fig.\ref{Fig_AzimuthalModelAnalytical} evaluates this model around the target order $\nu_{0}=N_{g}=228$, with
$n_{\mathrm{eff}}=3.2$, $\lambda=1310$~nm, and $R=14.86\,\mu$m chosen to satisfy the
resonance condition $2\pi R\,n_{\mathrm{eff}}=\nu_{0}\lambda$. The unperturbed quality
factor is estimated from evanescent leakage through the angular-momentum barrier,
\begin{equation}
    Q_{0}\approx\frac{1}{T},\qquad
    T=e^{-2\kappa_{\mathrm{ev}}w},\qquad
    \kappa_{\mathrm{ev}}=\sqrt{(\nu_{0}/R)^{2}-k_{0}^{2}}.
    \label{eq:Q0est}
\end{equation}

At the Bragg order the symmetric mode acquires the decay rate of
Eq.~\eqref{eq:gamma_g} with the prefactor set to unity, $\gamma_{g}^{S}/\omega_{r}=(d/R)^{2}\,
|f_{\nu}(R_{\mathrm{inner}})|^{2}$, using the fabricated grating depth $d=0.3\,\mu$m and
$|f_{\nu}(R_{\mathrm{inner}})|^{2}=0.25$ from the mode profile, giving
$Q_{S}\approx10^{4}$; the anti-symmetric mode has $\gamma_{g}^{A}=0$ by Eq.~\eqref{eq:gammaA}, so
$Q_{A}=Q_{0}$. Adjacent orders $\nu\neq N_{g}$ radiate through the travelling-wave
channel $\nu'=\nu-N_{g}$, which over the plotted range ($|\nu'|\le8$) lies deep inside
the light cone ($k_{0}R\approx71$); their loss is therefore taken equal to the resonant
coupling strength, scaled by a Lorentzian mismatch factor
$\bigl[1+(\Delta\nu/(N_{g}/10))^{2}\bigr]^{-1}\!\approx\!1$, which degrades both standing
waves equally and leaves the anti-symmetric mode at $\nu_{0}$ as the only high-$Q$
survivor.

\begin{figure}[htbp]
\centering\includegraphics[width=13cm]{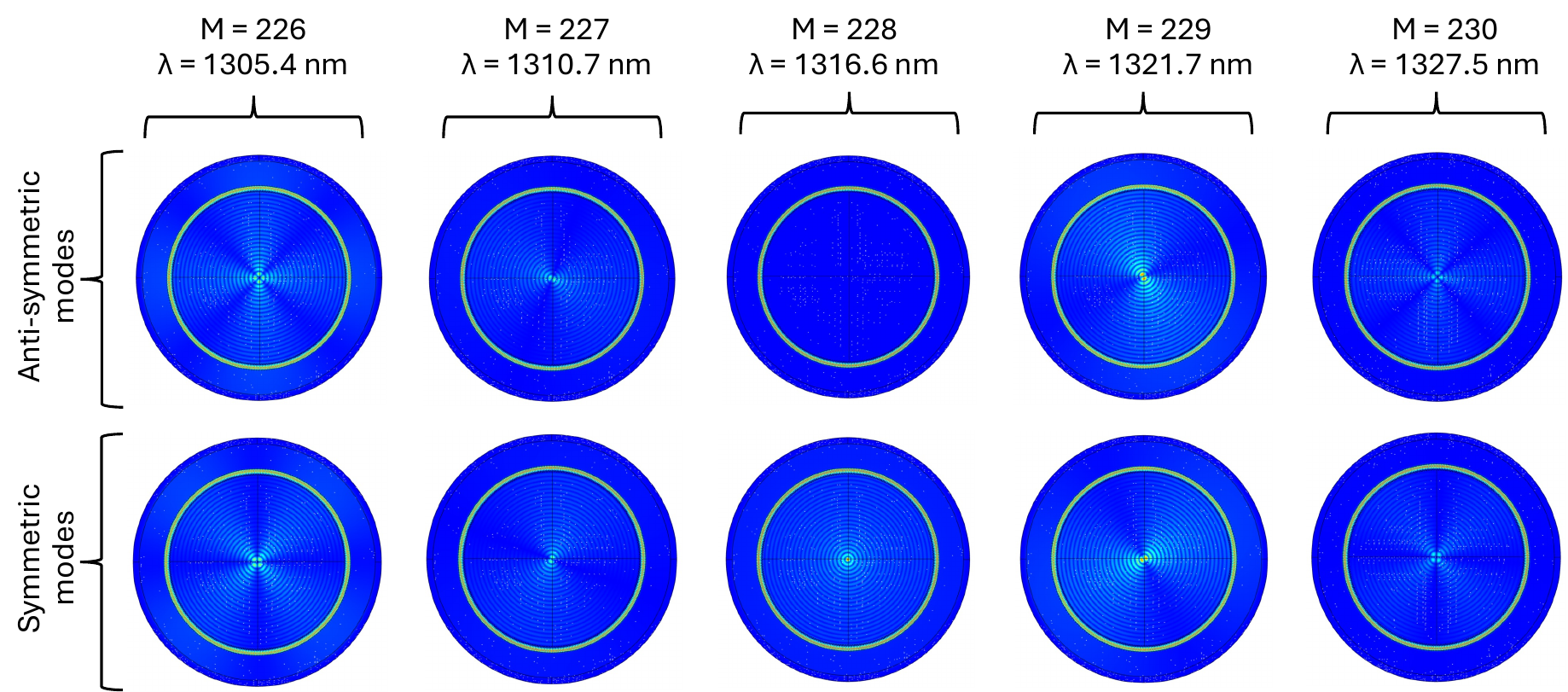}
\caption{Top-view of FEM simulated electric field for supported resonant modes with second order grating interaction according to $M \lambda /n_{eff} = 2 \pi R$. Favorable selective mode loss is observed for $M$ = 228.}
\label{Fig_SymmAsymmSims_v2}
\end{figure}

In order to validate the analytical model of selective mode loss for the 2$^{nd}$ order grating, a Finite Element Method (FEM) solver in COMSOL was used to perform numerical simulations on the quality factor (Q-factor) and loss ($\alpha_{i}$) of a target mode and its adjacent modes. The rectangular grating parameters are comprised of a silicon waveguide width $w_{si}$ = 1.0 $\mu$m, grating order $M$ = 228, fill-factor $FF$ = 0.50, and radius $R$ = 15 $\mu$m.  Fig. \ref{Fig_SymmAsymmSims_v2} presents the top-view simulated electric field distributions for both anti-symmetric and symmetric resonant modes across azimuthal mode numbers $M = 226$ - $230$ ($\lambda = 1305.4$ - $1327.5$ nm) in a silicon ring resonator satisfying the second-order grating Bragg condition $M\lambda/n_{\text{eff}} = 2\pi R$. At all mode numbers except $M = 228$, both anti-symmetric and symmetric modes exhibit pronounced radial field leakage toward the resonator center, manifesting as visible spoke-like intensity patterns that are indicative of significant grating-induced radiative loss coupling energy out of the whispering-gallery guided mode. 

\begin{figure}[htbp]
\centering\includegraphics[width=13.5cm]{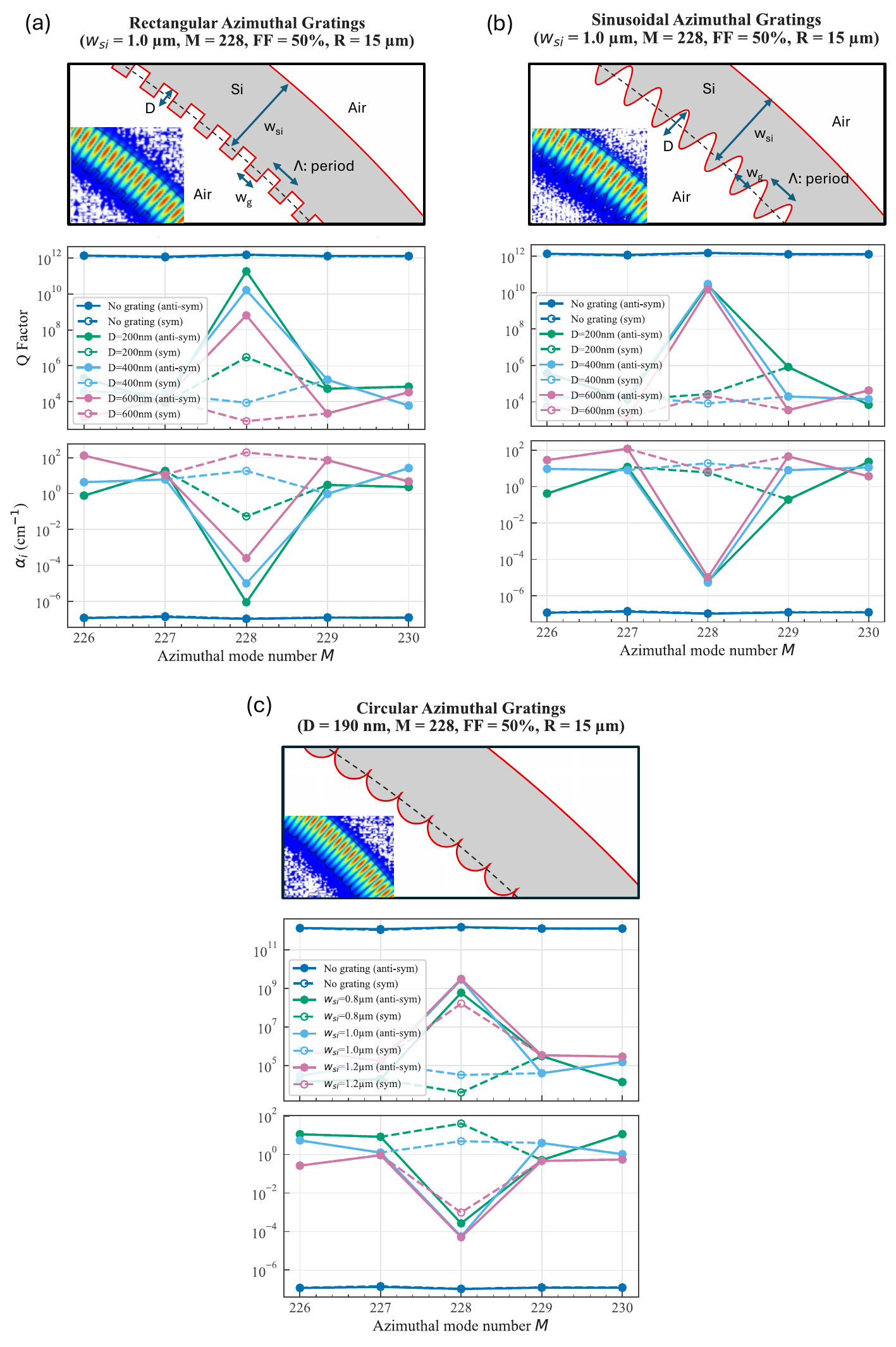}
\caption{Q-Factor and loss ($\alpha_i$) calculations of a ring resonator with silicon waveguide width $w_{si}$ = 1.0 $\mu$m, grating order $M$ = 228, fill-factor $FF$ = 0.50, radius $R$ = 15 $\mu$m for (a) rectangular and (b) sinusoidal azimuthal gratings. (c) Circular azimuthal grating performance for different silicon waveguide widths.}
\label{Fig_COMSOLsims_v3}
\end{figure}

In stark contrast, the anti-symmetric mode at $M = 228$ displays a notably clean, azimuthal uniform intensity envelope confined strictly within the annular waveguide with negligible central field intensity - demonstrating that this particular mode experiences near-zero radiative loss at the Bragg condition. The symmetric mode at $M = 228$, however, does not share this behavior; it retains strong radially directed emission toward the resonator center, confirming that the suppression of radiative loss at the Bragg resonance is exclusive to the anti-symmetric mode symmetry class. The most visually distinct behavior emerges at $M = 228$, where the symmetric mode shows markedly enhanced radial field leakage toward the resonator center, consistent with constructive interference of second-order diffracted fields satisfying the Bragg condition at this mode number. These field distributions directly corroborate the $Q$-factor and loss calculations of Fig. \ref{Fig_COMSOLsims_v3}a-c, confirming that selective modal loss are maximized at $M = 228$. It should also be noted that the symmetric mode at $M = 228$ shows strong directed emission toward the center, which while lossy, could be used to exploit intentional surface emission or vertical out-coupling rather than treated purely as a loss mechanism. Regarding wavelength selectivity, the approximately $5.3$ nm spacing between adjacent azimuthal mode numbers, spanning
$\lambda = 1305.4$ nm to $\lambda = 1327.5$ nm across four mode steps, defines the free spectral range (FSR) of the resonator. Notably, the selective loss discrimination between anti-symmetric and symmetric modes is operative over a very narrow wavelength window centered at $\lambda = 1316.6$ nm ($M = 228$), indicating that the grating-induced modal selectivity is highly wavelength-specific. This narrow operational bandwidth has direct implications for wavelength-division multiplexing (WDM) and optical sensing applications, where precise resonance control and sharp loss discrimination between adjacent channels are critical performance requirements.

Fig. \ref{Fig_COMSOLsims_v3}a-c presents the simulated $Q$-factor and propagation loss $\alpha_i$ as a function of azimuthal mode number $M$ for a silicon ring resonator ($w_{si} = 1.0~\mu$m, $R = 15~\mu$m, $FF = 0.50$) incorporating rectangular, sinusoidal, and circular azimuthal grating profiles. In the absence of any grating perturbation, both anti-symmetric and symmetric modes sustain $Q$-factors exceeding $10^{12}$ with propagation losses on the order of $10^{-6}$~cm$^{-1}$ across the entire mode spectrum $M = 226$--$230$, establishing the unperturbed resonator as 
the performance upper bound. Upon introduction of the azimuthal grating at the second-order Bragg condition ($M = 228$), all grating configurations produce a sharp, mode-selective resonance at $M = 228$, manifesting as a simultaneous maximum in $Q$-factor and minimum in $\alpha_i$, consistent with suppressed radiative out-coupling at the Bragg condition. Critically, at $M = 228$ the anti-symmetric mode exhibits a pronounced $Q$-factor enhancement relative to all off-resonance modes, while the 
symmetric mode is severely degraded relative to its anti-symmetric 
counterpart, confirming the symmetry-dependent nature of the grating--mode coupling discussed in relation to Fig. \ref{Fig_SymmAsymmSims_v2}.  For rectangular gratings 
(Fig. \ref{Fig_COMSOLsims_v3}a), increasing the grating depth $D$ from $200$~nm to $600$~nm monotonically degrades the $Q$-factor at $M = 228$ by several orders of magnitude --- from approximately $10^{11}$ at $D = 200$~nm down to $\sim$$10^{9}$ at $D = 600$~nm --- while leaving off-resonance modes comparatively unaffected, demonstrating that grating depth serves as the primary tuning parameter for loss engineering at the Bragg resonance. The sinusoidal grating holds the $Q$-factor near $10^{10}$ over the same range (Fig.~\ref{Fig_COMSOLsims_v3}b), attributed 
to the reduced higher-order Fourier harmonic content of the sinusoidal profile, which 
diminishes coupling to higher-order radiation channels. For circular azimuthal gratings 
(Fig.~\ref{Fig_COMSOLsims_v3}c), fixed at $D = 190$~nm and $M = 228$, varying the silicon waveguide width 
$w_{si}$ from $0.9~\mu$m to $1.2~\mu$m reveals a non-monotonic dependence of the 
$Q$-factor on waveguide geometry: narrower widths ($w_{si} = 0.9~\mu$m) produce the most 
severe loss peaks at $M = 228$, while intermediate widths ($w_{si} = 1.0$--$1.1~\mu$m) 
yield comparatively higher $Q$-factors, suggesting an optimal confinement regime in which 
the modal overlap with the grating perturbation is partially suppressed. Across all three 
grating geometries, the anti-symmetric mode consistently outperforms its symmetric counterpart 
in $Q$-factor retention at $M = 228$, reinforcing the conclusion that anti-symmetric modal 
symmetry is a necessary condition for achieving low radiative loss at the Bragg resonance. 
These results collectively establish a clear design framework: grating depth, profile 
geometry, and waveguide width can be co-optimized to achieve targeted $Q$-factors and 
emission rates in azimuthal grating-integrated ring resonators, with direct relevance to 
applications in narrow-linewidth lasing, optical sensing, and wavelength-selective 
add-drop filtering.

\section{Fabrication}
\begin{figure}[htbp]
\centering\includegraphics[width=13cm]{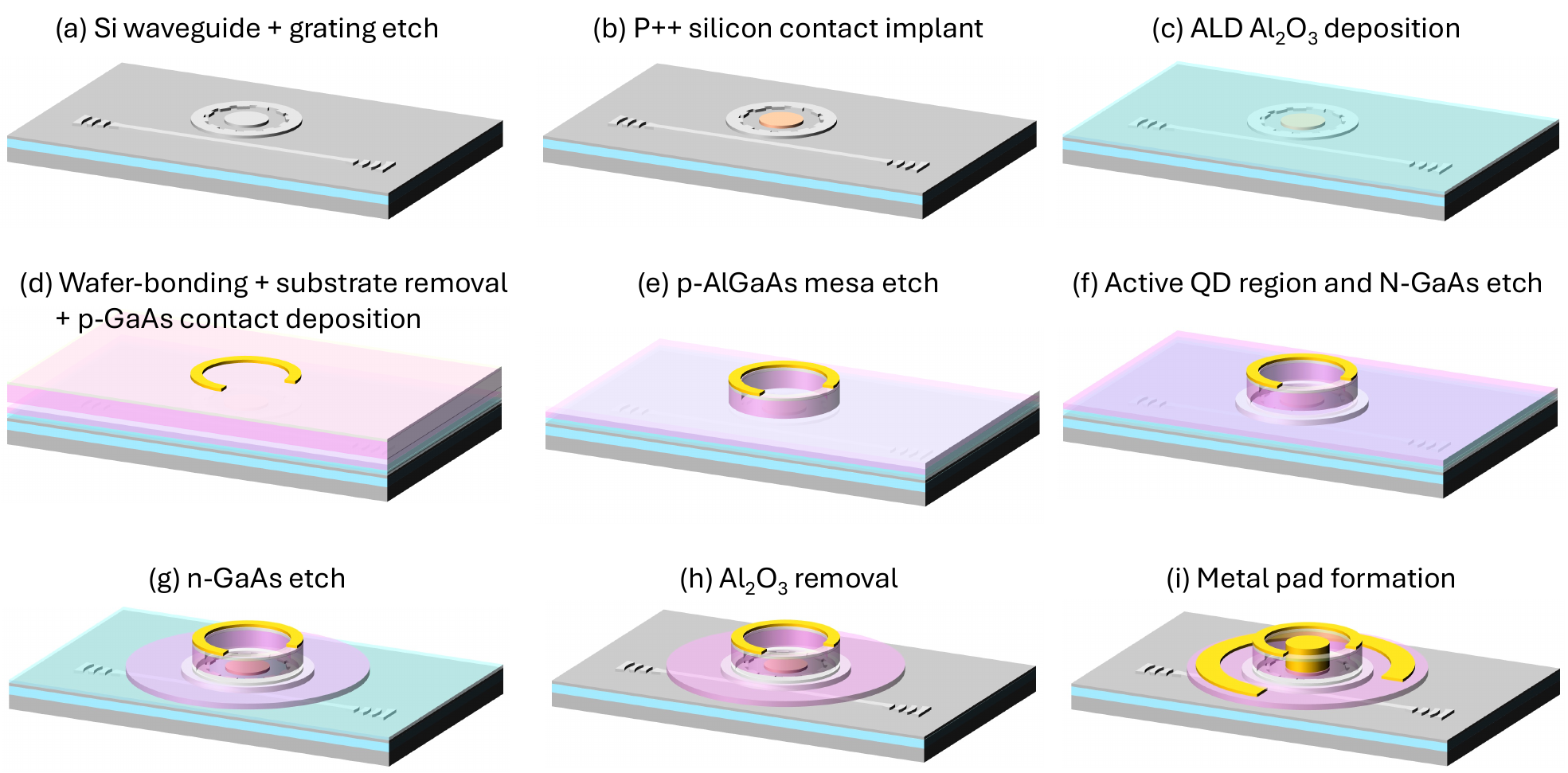}
\caption{Fabrication flow of III-V/Si micro-ring laser with azimuthal gratings.}
\label{Fig_FabricationSteps}
\end{figure}

Device fabrication was carried out in-house beginning with a 100 mm silicon-on-insulator (SOI) wafer comprised of a 300 nm silicon device layer atop a 2 $\mu$m buried oxide (BOX) layer. A blanket boron implantation ($4 \times 10^{16}$ cm$^{-3}$) was first ion implanted to provide electrical conductivity for the integrated volatile and non-volatile semiconductor-insulator-semiconductor capacitive (SISCAP) phase shifters. Alignment markers and grating couplers were patterned using a 248 nm KrF ASML deep ultraviolet (DUV) stepper and subsequently etched to a depth of 145 nm using Cl$_2$-based plasma chemistry as shown in Fig. \ref{Fig_FabricationSteps}a. To form the p$^{++}$ regions for the SISCAP devices, a nine-step boron implantation process was employed, achieving a doping concentration of $1 \times 10^{20}$ cm$^{-3}$ as shown in Fig. \ref{Fig_FabricationSteps}b. Silicon rib waveguides and vertical outgassing channels (VOCs) were then defined and etched to depths of 217 nm and 300 nm, respectively, with laser endpoint detection used to monitor the process.

Following silicon patterning, the wafer underwent an extensive cleaning sequence consisting of a Piranha clean, a brief dilute hydrofluoric acid dip to eliminate residual hard-mask material, an O$_2$ plasma treatment, and standard SC1/SC2 cleans. The III-V quantum-dot (QD) epitaxial wafer was separately cleaned using acetone, methanol, and isopropyl alcohol (IPA), followed by O$_2$ plasma exposure and a 1 min NH$_4$OH:H$_2$O (1:10) treatment. Subsequently, an Al$_2$O$_3$/HfO$_2$/Al$_2$O$_3$ dielectric stack was deposited onto both the III-V and SOI wafers using atomic layer deposition (ALD) at 300 $^\circ$C. The III-V and SOI wafers were manually aligned at room temperature using a Finetech flip-chip bonder and then bonded under pressure at 300 $^\circ$C using a 2-hour temperature ramp, with a total bonding duration of 15 hours. After bonding, the p-GaAs substrate was mechanically lapped until approximately 100 $\mu$m of material remained. The p-GaAs substrate was then selectively removed through wet chemical etching, terminating on a 20 nm p-AlGaAs etch-stop layer. Buffered HF was subsequently used to remove the etch-stop layer and expose a pristine 100 nm p-GaAs contact layer as indicated in Fig. \ref{Fig_FabricationSteps}d. Laser p-contact metallization consisting of Pt/Ti/Pt/Au (5/25/50/250 nm) was deposited onto the exposed p-GaAs surface. III-V mesas were then patterned using a SiN hard mask and etched via inductively coupled plasma (ICP) etching with Cl$_2$-based chemistry, stopping on a 100 nm n-AlGaAs etch-stop layer through laser endpoint monitoring. This remaining etch-stop layer was selectively removed using a wet etch to expose the n-GaAs contact layer as shown in Fig. \ref{Fig_FabricationSteps}f. The laser n-contact was subsequently formed using Pd/Ge/Ti/Au/Ti (30/60/50/200/10 nm) metallization followed by a rapid thermal anneal at 300 $^\circ$C for 30 seconds. To electrically isolate the III-V QD mesas, selected regions of the n-GaAs and ALD dielectric were dry etched. A PECVD SiN cladding layer was then deposited, followed by a thick benzocyclobutene (BCB) layer to reduce parasitic electrical effects. Finally, thick Ti/Au metal pads ($\sim$1.6 $\mu$m) were evaporated to provide probe contacts for both the p- and n-type laser electrodes.

\section{Characterization and Measurements}
 
\subsection{Measurement Preliminaries and Design of Experiment (DOE)}

Light--current--voltage (LIV) characteristics were measured under continuous-wave current injection using a Keithley 2400 source-measure unit, with the 100 mm wafer vacuum-mounted on a copper chuck held at $T = 25\ ^\circ$C by a closed-loop temperature controller. Optical power was collected from the left and right grating couplers using cleaved SMF-28 fibers positioned at an angle of $7^\circ$ and measured with a Newport 2936-C power meter. Optical spectra were acquired simultaneously by routing the collected light through a 99/1 directional coupler into a Yokogawa AQ6370E optical spectrum analyzer operated at a resolution bandwidth of 0.02 nm. Measured grating-coupler losses for TE polarization were determined to be on average $\sim -8.7$ dB/coupler at 1310 nm. All LI responses reported here are normalized to this value and quoted as the sum of the two facet outputs, i.e.\ as on-chip power delivered into the silicon bus waveguide. The bias-resolved spectral maps of Fig.~\ref{Fig_GratingVsNoGrating}c--d were assembled from current sweeps in 0.2 mA steps, with one full OSA trace recorded at every step. Threshold currents $I_{th}$ were extracted from the peak of the second derivative of the LI curve following Telcordia GR-468-CORE~\cite{Telcordia_GR468,ILXLightwave_AN12}, slope efficiencies $SE$ from a linear fit to the LI curve above threshold, and series resistances $R_{s}$ from a linear fit to the $I$--$V$ characteristic over 15--25 mA.
 
The purpose of this experiment is to establish the effect of the azimuthal grating on the lasing wavelength and the threshold current, and the stability of the wavelength control it provides. Two comparisons are made. The first isolates the grating: MRLs were patterned with and without the silicon inner-wall grating, with all other layers, dimensions and processing steps held fixed. The second changes only the ring-to-bus coupling angle, $\theta_{\mathrm{coupler}} = 0^\circ$, $40^\circ$ and $60^\circ$, leaving the ring itself unchanged; this sets the power coupling coefficient $k$ and with it the mirror loss, and tests whether the grating-defined wavelength remains fixed across different values of $k$.
 
All devices reported in this section share the same geometry: a ring radius of $R_{r} = 25\ \mu$m, a silicon waveguide width of $w_{si} = 1.0\ \mu$m, and a III--V mesa width of $w_{\mathrm{III\text{-}V}} = 5.0\ \mu$m. The azimuthal grating is of order $M = 373$ with a fill factor of $FF = 0.5$ and a grating depth of $0.3\ \mu$m.

\subsection{Impact of the Azimuthal Grating and Wavelength Reproducibility}
 
\begin{figure}[htbp]
\centering\includegraphics[width=14cm]{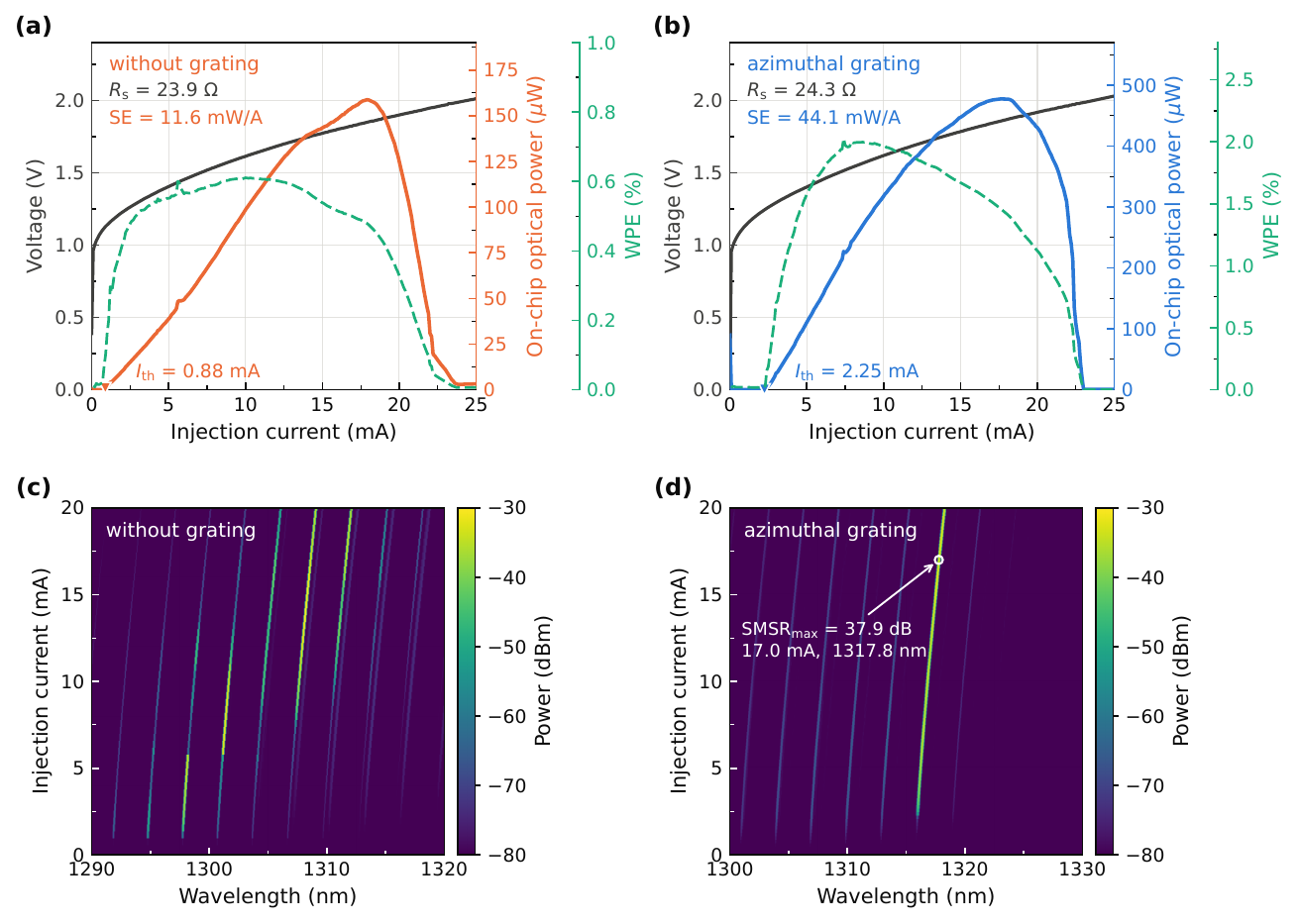}
\caption{Azimuthal-grating and grating-free hybrid III-V/Si QD micro-ring
lasers measured at $T = 25\,^{\circ}$C. (a),~(b) Light--current--voltage
characteristics of the grating-free reference and of the azimuthal-grating
device; the solid colored curves show the on-chip optical power summed over
both facets (inner right axis), the dashed curves show the corresponding
on-chip wall-plug efficiency (WPE, outer right axis), and the triangles mark
the extracted threshold currents. (c),~(d) Corresponding bias-resolved
spectral maps of the right-facet emission for the same two devices; the
circle in (d) marks the bias of maximum side-mode suppression.}
\label{Fig_GratingVsNoGrating}
\end{figure}

Fig.~\ref{Fig_GratingVsNoGrating} compares an MRL incorporating the azimuthal grating against a grating-free reference fabricated on the same die. The two devices are electrically identical: the $I$--$V$ characteristics plotted on the left axis of Fig.~\ref{Fig_GratingVsNoGrating}a--b overlap and yield series resistances of $R_{s} = 23.9\ \Omega$ and $24.3\ \Omega$, so the grating acts on the optical mode alone.

The grating raises the threshold current from $I_{th} = 0.88$ mA to $I_{th} = 2.25$ mA. Two mechanisms contribute. First, etching the corrugation into the inner sidewall of the silicon ring 
introduces additional sidewall roughness, and the resulting scattering loss adds to the round-trip cavity loss. Second, the grating-free reference lases at 1295--1301 nm at turn-on, at the gain peak itself, since without mode selection the longitudinal mode with the highest modal gain reaches threshold first. The grating fixes the operating wavelength at 1316.0 nm at turn-on, so the selected mode draws on the tail of the gain spectrum rather than on its maximum. 
 
The bias-resolved spectral maps of Fig.~\ref{Fig_GratingVsNoGrating}c--d resolve the mode selection directly. In the grating-free device (Fig.~\ref{Fig_GratingVsNoGrating}c) the entire whispering-gallery comb lases: multiple modes are lasing across the QD gain bandwidth, with a free spectral range of $\Delta\lambda_{\mathrm{FSR}} = 2.89$ nm at $\lambda = 1297.8$ nm, corresponding to a group index $n_{g} = \lambda^{2}/(\Delta\lambda_{\mathrm{FSR}}\cdot 2\pi R_{r}) = 3.71$. As the injection current is raised, the strongest line hops successively between adjacent comb orders, carrying the dominant lasing wavelength from 1295 nm to 1314 nm by 20 mA through a sequence of discrete jumps. Each individual comb line meanwhile red-shifts smoothly with bias, producing the characteristic set of parallel tilted stripes in the map.
 
With the azimuthal grating (Fig.~\ref{Fig_GratingVsNoGrating}d) the map collapses to a single trace. Lasing begins at 1316.0 nm just above 2.25 mA and remains on the same longitudinal order over the entire bias range, reaching 1317.5 nm at 15 mA. A fit to the eight consecutive comb orders resolved between 1299.7 nm and 1320.5 nm gives $\Delta\lambda_{\mathrm{FSR}} = 2.976$ nm and \ $n_{g} = 3.71$. The SMSR rises from $\sim 20$ dB at turn-on to 37.9 dB at 17.0 mA and 1317.8 nm, marked in Fig.~\ref{Fig_GratingVsNoGrating}d, and remains above 30 dB from 8.2 mA to 20 mA with no mode hops. The selected line red-shifts at 0.129 nm/mA, set by junction self-heating acting through the thermo-optic coefficient of the hybrid mode. This continuous, hop-free tuning is what makes the device usable as a DWDM channel source: the emission wavelength can be trimmed by bias without the discontinuous jumps of the grating-free ring.

\begin{figure}[htbp]
\centering\includegraphics[width=13cm]{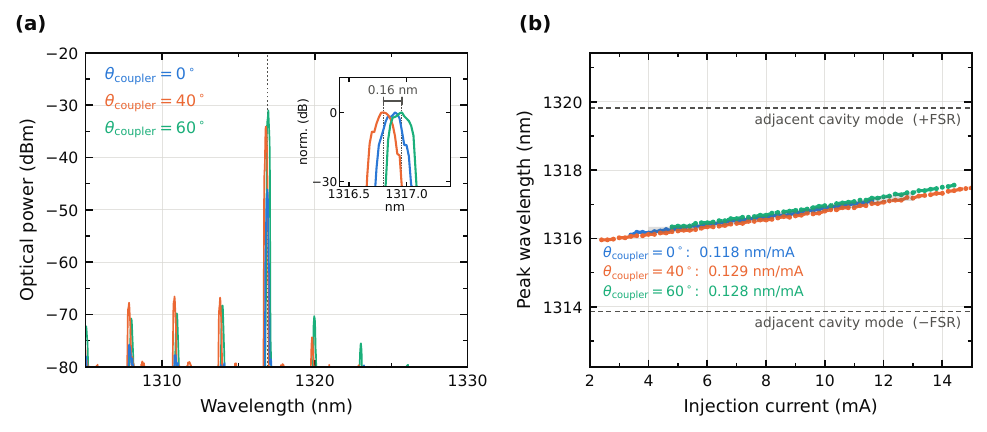}
\caption{Wavelength reproducibility of azimuthal-grating MRLs across three ring-to-bus coupling-angle designs ($\theta_{\mathrm{coupler}} = 0^\circ$, $40^\circ$, $60^\circ$) sharing the same grating and ring geometry, at $T = 25\ ^\circ$C. (a) Optical spectra at a common injection current of 10 mA; the dotted line marks the mean lasing wavelength and the inset shows the same three lines normalized to their own peaks. (b) Peak lasing wavelength versus injection current over each device's single-mode operating range; the dashed lines mark the adjacent longitudinal modes, and the annotated slopes are the thermal tuning rates.}
\label{Fig_WavelengthLock}
\end{figure}

Fig.~\ref{Fig_WavelengthLock} examines whether the emission wavelength is set by the grating or simply by whichever longitudinal mode of the ring cavity lies closest to the gain peak. Three azimuthal-grating MRLs are compared that share the same grating design and ring geometry but differ in the wrapped directional-coupler angle, $\theta_{\mathrm{coupler}} = 0^\circ$, $40^\circ$ and $60^\circ$, and therefore in the power coupling coefficient $k$ and the mirror loss $\alpha_{m}$ of the resonant mode.
 
The three designs differ substantially as lasers, with threshold currents of 3.39, 2.25 and 4.71 mA for $\theta_{\mathrm{coupler}} = 0^\circ$, $40^\circ$ and $60^\circ$ respectively, as expected from the different mirror losses set by the coupler angle. Output powers, slope efficiencies and the remaining extracted parameters for all devices in this section are summarized in Table~\ref{tab:measured_devices}.
 
The spectra of Fig.~\ref{Fig_WavelengthLock}a, taken at a common bias of 10 mA, place all three devices on the same wavelength. They lase on a single mode at 1316.90, 1316.80 and 1316.96 nm, a total spread of 0.16 nm, with SMSRs of 29.6, 32.5 and 37.2 dB. The residual longitudinal mode comb appears in each trace at the same 2.98 nm spacing, and the same comb order is selected in every case. The inset, in which each trace is normalized to its own peak, resolves the three lines.
 
Fig.~\ref{Fig_WavelengthLock}b extends the comparison across bias on a wavelength axis spanning $\pm 1.5$ free spectral ranges, so that the spread is read against the natural wavelength scale of the cavity. Over the 4--13 mA range in which all three devices operate single-mode, the peak wavelength of each rises linearly with current and the three curves remain parallel: the maximum device-to-device spread is 0.27 nm, less than one tenth of the 2.98 nm mode spacing, and the thermal tuning rates agree to within 8\% (0.118, 0.129 and 0.128 nm/mA).
 
Changing the coupler angle from $0^\circ$ to $60^\circ$ changes $k$ and with it the mirror loss $\alpha_{m}$, the threshold carrier density and the intracavity photon density, varying the threshold current by more than a factor of two across the set. The lasing wavelength moves by less than a tenth of the mode spacing across that range, and the tuning rate is common to all three designs. The operating wavelength is therefore pinned by the lithographically defined grating period, consistent with the symmetry-selective loss mechanism of Sec.~2.2, with the common tuning rate the signature of junction self-heating acting on a fixed, grating-defined resonance.
 
Table~\ref{tab:measured_devices} extends the comparison to grating-free rings fabricated at all three coupler angles. Their thresholds are lower, 1.27, 0.88 and 1.25 mA, since they lase at the gain peak and carry no grating-induced loss, but none of them holds a wavelength. As the bias is swept from turn-on to 20 mA the lasing line of each migrates across a wide band of the QD gain spectrum through successive mode hops, covering 1283--1302 nm at $\theta_{\mathrm{coupler}} = 0^\circ$, 1295--1314 nm at $40^\circ$ and 1287--1322 nm at $60^\circ$: spans of 19, 19 and 35 nm, or six to twelve free spectral ranges. The three bands are offset from one another by more than a mode spacing, so the emission wavelength of a grating-free ring is neither fixed by bias nor reproducible from device to device.

\begin{table}[htbp]
\caption{Measured performance of the azimuthal-grating and grating-free III-V/Si QD MRLs at $T = 25\ ^\circ$C.}
\centering
\fontsize{8pt}{9.5pt}\selectfont
\setlength{\tabcolsep}{4.5pt}
\begin{tabular}{l c c c c c c}
\hline
 & \multicolumn{3}{c}{\textbf{No grating}} & \multicolumn{3}{c}{\textbf{Azimuthal grating}} \\
\cline{2-4}\cline{5-7}
$\theta_{\mathrm{coupler}}$ & $0^\circ$ & $40^\circ$ & $60^\circ$ & $0^\circ$ & $40^\circ$ & $60^\circ$ \\
\hline
$I_{th}$ [mA]                       & 1.27 & 0.88 & 1.25 & 3.39 & 2.25 & 4.71 \\
$SE$ [mW/A]                         & 4.8 & 11.6  & 23.9 & 4.7  & 44.1 & 48.2 \\
Peak power [$\mu$W]         & 68  & 159   & 201  & 32   & 478  & 359  \\
Peak WPE [\%]                       & 0.42 & 0.61 & 1.67 & 0.19 & 2.03 & 1.70 \\
$R_{s}$ [$\Omega$]                  & 22.4 & 23.9 & 21.9 & 25.0 & 24.3 & 25.7 \\
$\lambda$ @ 10 mA [nm]              & multi-mode & multi-mode & multi-mode & 1316.90 & 1316.80 & 1316.96 \\
Max SMSR [dB]                       & --   & --   & --   & 29.6 & 37.9 & 37.2 \\
Lasing $\lambda$ range [nm]         & 1283--1302 & 1295--1314 & 1287--1322 & 1316.2--1317.4 & 1316.0--1318.3 & 1316.2--1317.6 \\
\hline
\end{tabular}
\label{tab:measured_devices}
\end{table}
 
With the azimuthal grating the same three coupler designs confine the emission to 1316.2--1317.4, 1316.0--1318.3 and 1316.2--1317.6 nm over the same bias sweep, spans of 1.2 to 2.3 nm that lie within a single free spectral range of one another and are traversed continuously by thermal tuning rather than by mode hops. The accessible wavelength band is narrowed by more than an order of magnitude, and a bias- and device-dependent emission wavelength becomes a fixed, lithographically addressed one.
Channel placement is consequently set at mask level and is robust to the coupler design chosen for a given power or efficiency target. The device-to-device spread is an order of magnitude smaller than the cavity mode spacing and can be absorbed by a small bias adjustment, while the common thermal tuning rate means that a single calibration law applies across the die. These two properties---lithographically defined emission wavelengths and a shared, predictable tuning response---are precisely what a multi-wavelength DWDM source requires of its constituent lasers.

\subsection{Temperature Stability and Gain Detuning}
 
\begin{figure}[htbp]
\centering\includegraphics[width=12cm]{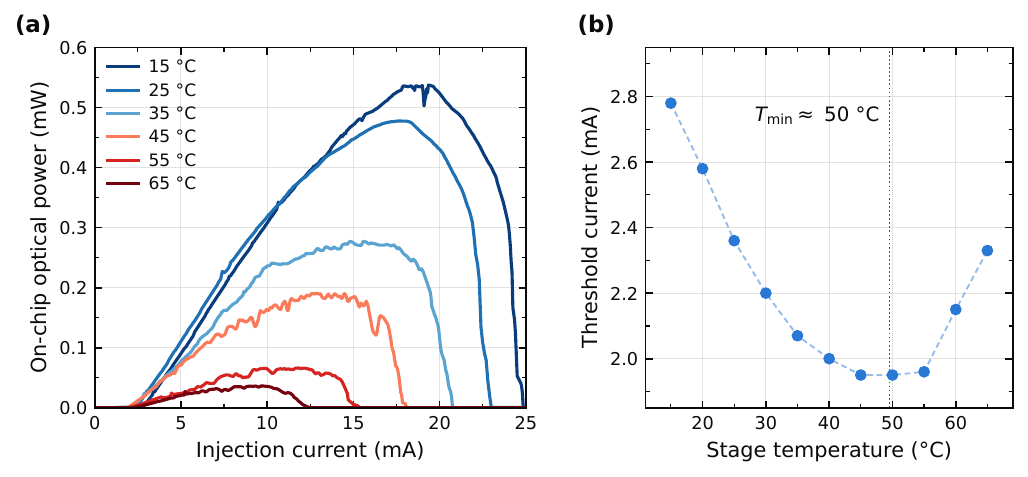}
\caption{Temperature dependence of the azimuthal-grating MRL ($\theta_{\mathrm{coupler}} = 40^\circ$) over 15--65\ $^\circ$C. (a) On-chip LI characteristics at selected stage temperatures. (b) Extracted threshold current versus stage temperature; the dotted line marks the threshold minimum.}
\label{Fig_LIV_T}
\end{figure}

The azimuthal-grating MRL with $\theta_{\mathrm{coupler}} = 40^\circ$ was characterized over a stage temperature range of 15--65\ $^\circ$C in 5\ $^\circ$C steps as shown in Fig. \ref{Fig_LIV_T}a. At each temperature the LIV characteristic was recorded and a full optical spectrum was acquired at a fixed bias of 5.0 mA. The grating-free reference ring of Section 4.2 was measured over the same temperature range at a bias just above its threshold, and the peak of its whispering-gallery envelope was used to track the QD gain peak.
 
The threshold current does not increase monotonically with temperature. Fig.~\ref{Fig_LIV_T}b shows $I_{th}$ falling from 2.78 mA at 15\ $^\circ$C to a minimum of 1.95 mA between 45 and 50\ $^\circ$C, a reduction of 30\%, before rising to 2.33 mA at 65\ $^\circ$C. Fitted to $I_{th} \propto \exp(T/T_{0})$, the two branches give characteristic temperatures of $T_{0} = -82$ K over 15--45\ $^\circ$C and $T_{0} = +80$ K over 50--65\ $^\circ$C. Every other figure of merit degrades monotonically over the same range, as the LI family of Fig.~\ref{Fig_LIV_T}a shows: the slope efficiency falls by a factor of 5.6, the peak on-chip power drops from 0.54 mW to 0.04 mW, the roll-over current moves from 19.4 mA to 9.7 mA, and the series resistance rises from 23.5 to 25.9 $\Omega$. The threshold minimum is therefore not a general improvement in device performance but the signature of a mechanism that acts specifically on the lasing wavelength.
 
\begin{figure}[htbp]
\centering\includegraphics[width=12cm]{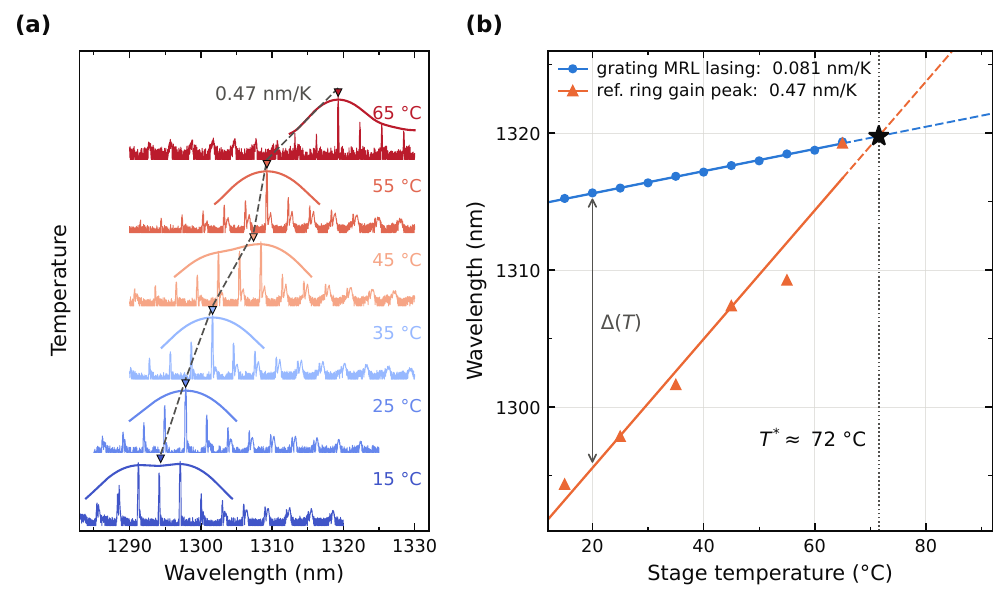}
\caption{Thermal decoupling of the grating-defined wavelength from the QD gain peak. (a) Spectra of the grating-free reference ring at each stage temperature, offset vertically, with the fitted gain envelope and its peak (triangles). (b) Lasing wavelength of the grating MRL and gain peak of the reference ring versus stage temperature; the star marks the extrapolated crossing $T^{*}$ and $\Delta(T)$ is the detuning.}
\label{Fig_detuning}
\end{figure}

Fig.~\ref{Fig_detuning} identifies that mechanism. The grating-defined emission wavelength shifts linearly from 1315.48 nm at 15\ $^\circ$C to 1319.56 nm at 65\ $^\circ$C, a rate of $d\lambda/dT = 0.081$ nm/K. The device remains on the same longitudinal order across the entire 50 K range, with an SMSR of 31.9 dB at 15\ $^\circ$C, above 30 dB up to 55\ $^\circ$C and 27.1 dB at 65\ $^\circ$C. The gain peak of the reference ring, shown as the spectral envelopes of Fig.~\ref{Fig_detuning}a, moves at 0.47 nm/K over the same range. Because the two shift at different rates, the detuning $\Delta(T)$ between the grating-defined wavelength and the gain peak is a function of temperature: it is 18 nm at 25\ $^\circ$C and closes at 0.39 nm/K, leaving 3 nm at 65\ $^\circ$C, with the extrapolated crossing $\Delta = 0$ at $T^{*} \approx 72\ ^\circ$C marked in Fig.~\ref{Fig_detuning}b. This accounts for the non-monotonic threshold. Between 15 and 50\ $^\circ$C the closing detuning raises the modal gain available at the fixed lasing wavelength faster than carrier leakage and gain-peak broadening raise the transparency current, and $I_{th}$ falls; above 50\ $^\circ$C the intrinsic degradation of the QD gain dominates and $I_{th}$ turns around. The minimum sits some 20 K below $T^{*}$, where the two contributions balance.
 
Two consequences follow. First, the 5.8-fold weaker temperature dependence
of the emission wavelength relaxes the temperature stabilization required
to hold the lasing wavelength, an attractive property for co-packaged
optics, where the cooling budget per transmitter is limited. Second, the
detuning $\Delta$ becomes a design variable in its own right: the grating
fixes $\lambda$ lithographically while the gain peak is set by the epitaxy
and moves with temperature, so the room-temperature offset between the
two---and with it the temperature of minimum threshold current---is chosen
at mask level. The 18~nm of detuning implemented here places that minimum
near 50~$^{\circ}$C, representative of junction temperatures under load;
more generally, a larger room-temperature detuning shifts this optimum to
higher temperature, so that the device reaches its smallest detuning---and
lowest threshold---precisely where the transmitter runs hot. A grating-free
ring offers no such degree of freedom: its lasing wavelength is slaved to
the gain peak, and its detuning vanishes at every temperature by
construction.

\subsection{High-Speed Measurements}
 
High-speed characterization was performed on the azimuthal-grating MRL with
$\theta_{\mathrm{coupler}} = 40^{\circ}$ from Secs.~4.2--4.3. The device was
directly probed using a signal--ground (SG) radio-frequency (RF) probe, and the
small-signal response $S_{21}$ was measured with an Agilent E8364B vector
network analyzer from 10~MHz to 20~GHz. The stage temperature was held at
$T = 25\,^{\circ}$C and the injection current was varied from 3 to 17~mA.

\begin{figure}[ht!]
    \centering
    \includegraphics[width=13cm]{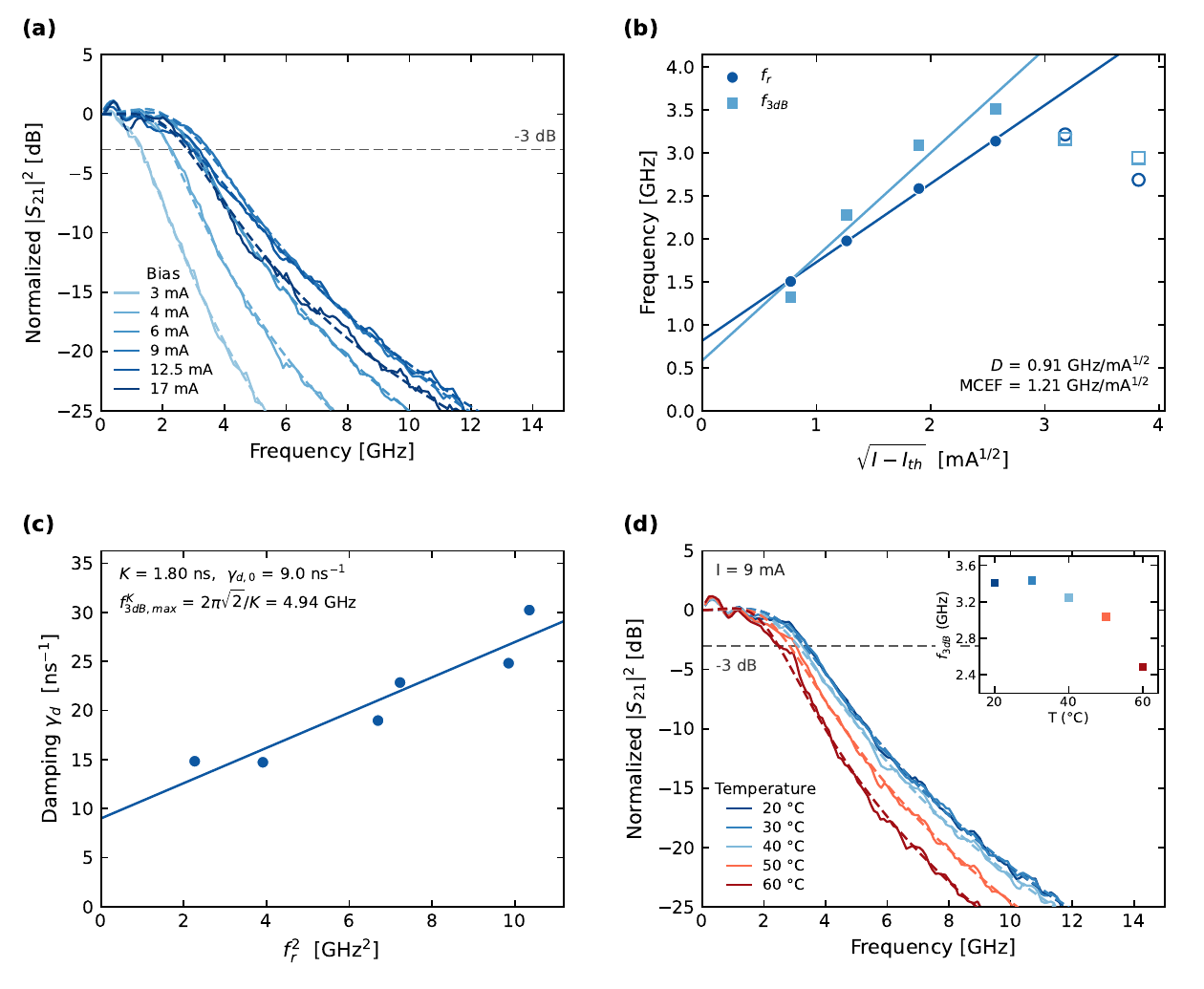}
    \caption{High-speed small-signal modulation characteristics of the
    azimuthal-grating MRL ($\theta_{\mathrm{coupler}} = 40^{\circ}$).
    (a)~Normalized $|S_{21}|^2$ responses (solid lines) measured at
    $T = 25\,^{\circ}$C for bias currents of 3--17~mA, together with the
    three-pole fits (dashed lines).
    (b)~$f_r$ (circles) and $f_{3\,\mathrm{dB}}$ (squares) versus
    $\sqrt{I - I_{th}}$ ($I_{th} = 2.25$~mA) with linear fits yielding
    MCEF~=~1.21~GHz/mA$^{1/2}$ and $D$~=~0.91~GHz/mA$^{1/2}$; open symbols
    mark the thermally saturated bias points (12.5 and 17~mA) excluded from
    the fits.
    (c)~Damping rate $\gamma_d$ versus $f_r^2$ with the linear fit of
    Eq.~(\ref{eq:kfactor}), yielding $K = 1.80$~ns and
    $\gamma_{d,0} = 9.0$~ns$^{-1}$, corresponding to a $K$-factor-limited
    bandwidth ceiling of $2\pi\sqrt{2}/K = 4.94$~GHz.
    (d)~Normalized $|S_{21}|^2$ responses (solid lines) and fits (dashed
    lines) at a fixed bias of 9~mA for stage temperatures of
    20--60~$^{\circ}$C; inset: extracted $f_{3\,\mathrm{dB}}$ versus stage
    temperature, which is flat between 20 and 30~$^{\circ}$C and degrades
    rapidly above 50~$^{\circ}$C.}
    \label{fig:highspeed}
\end{figure}
 
Fig.~\ref{fig:highspeed}(a) shows the measured $S_{21}$ responses at each
bias, normalized at low frequency, where a maximum 3-dB bandwidth of 3.5~GHz is
attained at a bias current of 9~mA. The responses are further fit using a
three-pole fitting function $H(f)$~\cite{Nagarajan1992_JQE}, presented as the dashed
lines in Fig.~\ref{fig:highspeed}(a), from which the damping rate $\gamma_d$
and relaxation oscillation frequency $f_r$ at each bias current are extracted.
The extracted $f_r$, together with the measured $f_{3\,\mathrm{dB}}$, are
plotted in Fig.~\ref{fig:highspeed}(b) as a function of the square root of the
bias current above threshold. Modulation efficiencies of 1.21~GHz/mA$^{1/2}$
for $f_{3\,\mathrm{dB}}$ (MCEF) and 0.91~GHz/mA$^{1/2}$ for $f_r$
($D$-factor) are extracted by linear fitting, using the data points below
$(I_b - I_{th})^{1/2} = 2.6$~mA$^{1/2}$ ($I_{th} = 2.25$~mA); the two highest
bias points deviate from the linear trend due to thermal saturation (open
symbols) and are excluded from the fits. By linearly fitting the damping rate
$\gamma_d$ versus squared $f_r$, shown in Fig.~\ref{fig:highspeed}(c), the
$K$-factor is 1.80~ns, and a maximum $K$-factor-limited
$f_{3\,\mathrm{dB,max}} = 2\pi\sqrt{2}/K$ of 4.94~GHz is calculated using the
equation~\cite{Inoue2018_OE}:
\begin{equation}
    \gamma_d = K \cdot f_r^{2} + \gamma_{d,0},
    \label{eq:kfactor}
\end{equation}
where $\gamma_{d,0}$ represents the damping offset (9.0~ns$^{-1}$). The
maximum measured $f_{3\,\mathrm{dB}}$ of 3.5~GHz is lower than the
4.3--5.0~GHz attained by grating-free hybrid QD MRLs on the same
platform~\cite{yang2026_experimentaldesign}. This reduction is attributed to the additional
loss and the gain detuning introduced by the azimuthal grating: the former
raises the threshold gain ($I_{th}$ increases from 0.88 to 2.25~mA in
Sec.~4.2), pushing the operating point toward QD gain saturation and lowering
the differential gain, while the latter places the lasing wavelength away from
the gain peak, with a detuning of $\sim$18~nm at 25~$^{\circ}$C, further reducing the differential gain available
at the lasing wavelength.
 
To examine the influence of temperature, the stage temperature was increased
from 20 to 60~$^{\circ}$C at a fixed bias of 9~mA
[Fig.~\ref{fig:highspeed}(d)]. As shown in the inset, $f_{3\,\mathrm{dB}}$
remains flat at $\sim$3.4~GHz between 20 and 30~$^{\circ}$C, decreases slowly
thereafter, and drops rapidly above 50~$^{\circ}$C, reaching 2.5~GHz at
60~$^{\circ}$C. The flatness at the low-temperature end can be explained by
the gain detuning discussed in Sec.~4.3: the detuning between the
grating-defined wavelength and the gain peak grows as the temperature
decreases, offsetting the intrinsically higher
differential gain at lower temperatures, so that cooling the device does not
appreciably increase the bandwidth, while above 50~$^{\circ}$C the degradation
is dominated by the intrinsic thermal degradation of the QD gain. The device
maintains $f_{3\,\mathrm{dB}} \geq 3$~GHz up to 50~$^{\circ}$C without
temperature control.

\section{Conclusion}
 
We have patterned azimuthal gratings into the silicon ring of a hybrid III--V/Si InAs/GaAs quantum-dot micro-ring laser and shown that a single lithographic degree of freedom converts a multi-longitudinal-mode whispering-gallery cavity into a wavelength-addressed, single-mode O-band source.
 
On the design side, the coupled-mode analysis of Sec.~2.2 derives the angular momentum selection rule $\nu' = \nu + qN_{g}$ from first principles and shows that the inner-wall corrugation replaces the degenerate counter-propagating pair by symmetric and anti-symmetric standing-wave supermodes with unequal grating-induced decay rates. At the second-order Bragg condition $N_{g} = \nu$ the overlap of the anti-symmetric mode with the dominant radiation channel vanishes identically, $\gamma_{g}^{A} = 0$, while that of the symmetric mode is maximal, so a symmetry-protected high-$Q$ state exists at exactly one azimuthal order. Finite-element simulations reproduce this at $M = 228$ ($\lambda = 1316.6$ nm) and identify grating depth as the primary loss-engineering handle, degrading the $Q$-factor at the Bragg order from $\sim 10^{11}$ at $D = 200$ nm to $\sim 10^{9}$ at $D = 600$ nm while leaving off-resonance orders unaffected.
 
Experimentally, devices fabricated in-house on a 100 mm SOI platform confirm the mechanism. Against a grating-free reference on the same die, whose dominant line hops through more than four free spectral ranges between 1295 and 1314 nm over a 20 mA sweep, the grating device holds a single longitudinal order from 1316.0 to 1317.5 nm with a maximum SMSR of 37.9 dB, suppression above 30 dB from 5 to 20 mA without mode hops, and continuous tuning at 0.129 nm/mA; the threshold current rises from 0.88 to 2.25 mA. Across three ring-to-bus coupling-angle designs that vary the threshold current, the lasing wavelength moves by no more than 0.27 nm (47 GHz) over the shared 4--13 mA range, less than one tenth of the 2.98 nm mode spacing, and the thermal tuning rate is common to all three. The emission wavelength is set by the lithographically defined grating period rather than by where a whispering-gallery resonance falls relative to the gain peak.
 
Under thermal load, the same decoupling holds. Over 15--65\ $^\circ$C the grating-defined wavelength shifts at 0.081 nm/K against 0.47 nm/K for the gain peak of the reference ring, a suppression of 5.8$\times$, with the device remaining on the same longitudinal order and above 27 dB SMSR throughout. The detuning between the two is therefore temperature dependent, 18 nm at 25\ $^\circ$C closing at 0.39 nm/K toward $T^{*} \approx 72\ ^\circ$C, and it drives a non-monotonic threshold current with a minimum of 1.95 mA near 50\ $^\circ$C. Detuning becomes a design variable in its own right: the room-temperature offset between grating and gain peak is chosen at mask level, and with it the temperature at which the threshold is minimized. Small-signal characterization gives a maximum 3-dB bandwidth of 3.5 GHz at 9 mA against a $K$-factor-limited ceiling of 4.94 GHz, remaining above 3 GHz up to 50\ $^\circ$C without temperature control; the reduction relative to grating-free QD MRLs follows from the same added loss and gain detuning that raise the threshold, both accessible through the grating depth and the designed detuning.
 
Collectively, these results establish the azimuthal grating as a practical route to wavelength control in hybrid III--V/Si QD micro-ring lasers. A single lithographic feature fixes the emission wavelength and holds it against a factor-of-two change in cavity loading and a 50 K temperature excursion, with no additional epitaxy, regrowth or post-fabrication trimming. Combined with milliamp-scale thresholds, side-mode suppression beyond 37 dB and multi-gigahertz direct modulation, this positions azimuthal-grating MRLs as building blocks for cascaded, wavelength-addressed transmitter arrays in DWDM data communication and co-packaged optics.

\begin{backmatter}
 
\bmsection{Funding}
Advanced Research Projects Agency-Energy (DE-AR0001039).
 
\bmsection{Acknowledgment}
We thank Hewlett Packard Labs and the UCSB nanofabrication facilities.
 
\bmsection{Disclosures}
The authors declare no conflicts of interest.
 
\bmsection{Data availability}
The data that support the findings of this study are not publicly available at this time but may be obtained from the authors upon reasonable request.
 
\end{backmatter}

\bibliography{sample}

\begin{thebibliography}{10}
\newcommand{\enquote}[1]{``#1''}

\bibitem{Liang2016_NP}
D.~Liang, X.~Huang, G.~Kurczveil, \emph{et~al.}, \enquote{Integrated finely tunable microring laser on silicon,} {\protect\JournalTitle{Nat. Photonics}} \textbf{10}, 719--722 (2016).

\bibitem{Cheung2025_NC}
S.~Cheung, Y.~London, Y.~Yuan, \emph{et~al.}, \enquote{Heterogeneous {III-V/Si} micro-ring laser array with multi-state non-volatile memory for ternary content-addressable memories,} {\protect\JournalTitle{Nat. Commun.}} \textbf{16}, 5020 (2025).

\bibitem{Zhang2019_Optica}
C.~Zhang, D.~Liang, G.~Kurczveil, \emph{et~al.}, \enquote{Hybrid quantum-dot microring laser on silicon,} {\protect\JournalTitle{Optica}} \textbf{6}, 1145--1151 (2019).

\bibitem{Wan_2018_ACS}
Y.~Wan, D.~Jung, C.~Shang, \emph{et~al.}, \enquote{Low-threshold continuous-wave operation of electrically pumped 1.55 $\mu$m {InAs} quantum dash microring lasers,} {\protect\JournalTitle{ACS Photonics}} \textbf{6}, 279--285 (2018).

\bibitem{Wan_Optica_2017}
Y.~Wan, J.~Norman, Q.~Li, \emph{et~al.}, \enquote{1.3 $\mu$m submilliamp threshold quantum dot micro-lasers on {Si},} {\protect\JournalTitle{Optica}} \textbf{4}, 940--944 (2017).

\bibitem{Wan_PR_2018}
Y.~Wan, D.~Inoue, D.~Jung, \emph{et~al.}, \enquote{Directly modulated quantum dot lasers on silicon with a milliampere threshold and high temperature stability,} {\protect\JournalTitle{Photon. Res.}} \textbf{6}, 776--781 (2018).

\bibitem{Liang_JSTQE_2011}
D.~Liang, M.~Fiorentino, S.~Srinivasan, \emph{et~al.}, \enquote{Low threshold electrically-pumped hybrid silicon microring lasers,} {\protect\JournalTitle{IEEE J. Sel. Top. Quantum Electron.}} \textbf{17}, 1528--1533 (2011).

\bibitem{Zhang_JSTQE_2011}
C.~Zhang, D.~Liang, G.~Kurczveil, \emph{et~al.}, \enquote{Thermal management of hybrid silicon ring lasers for high temperature operation,} {\protect\JournalTitle{IEEE J. Sel. Top. Quantum Electron.}} \textbf{21}, 385--391 (2015).

\bibitem{Liang_PJ_2011}
D.~Liang, M.~Fiorentino, S.~Srinivasan, \emph{et~al.}, \enquote{Optimization of hybrid silicon microring lasers,} {\protect\JournalTitle{IEEE Photonics J.}} \textbf{3}, 580--587 (2011).

\bibitem{Cheung_ISLC_2024}
S.~Cheung, D.~Liang, Y.~Yuan, \emph{et~al.}, \enquote{Hybrid {III-V/Si} micro-ring laser with non-volatile charge trap memory,} in \emph{2024 IEEE 29th International Semiconductor Laser Conference (ISLC),}  (2024), pp. 1--2.

\bibitem{Cheung_OFC_2025}
S.~Cheung, Y.~London, Y.~Yuan, \emph{et~al.}, \enquote{Arrays of non-volatile {III-V/Si} micro-ring lasers for memory search applications,} in \emph{2025 Optical Fiber Communications Conference and Exhibition (OFC),}  (2025), pp. 1--3.

\bibitem{Spuesens2011_G4}
T.~Spuesens, D.~Van~Thourhout, P.~Rojo-Romeo, \emph{et~al.}, \enquote{{CW} operation of {III--V} microdisk lasers on {SOI} fabricated in a 200 mm {CMOS} pilot line,} in \emph{8th IEEE International Conference on Group IV Photonics,}  (2011), pp. 199--201.

\bibitem{Sui2015_PR}
S.-S. Sui, M.-Y. Tang, Y.-D. Yang, \emph{et~al.}, \enquote{Investigation of hybrid microring lasers adhesively bonded on silicon wafer,} {\protect\JournalTitle{Photon. Res.}} \textbf{3}, 289--295 (2015).

\bibitem{Campenhout2007_OE}
J.~V. Campenhout, P.~Rojo-Romeo, P.~Regreny, \emph{et~al.}, \enquote{Electrically pumped {InP}-based microdisk lasers integrated with a nanophotonic silicon-on-insulator waveguide circuit,} {\protect\JournalTitle{Opt. Express}} \textbf{15}, 6744--6749 (2007).

\bibitem{Campenhout2008_PTL}
J.~Van~Campenhout, L.~Liu, P.~Rojo~Romeo, \emph{et~al.}, \enquote{A compact {SOI}-integrated multiwavelength laser source based on cascaded {InP} microdisks,} {\protect\JournalTitle{IEEE Photonics Technol. Lett.}} \textbf{20}, 1345--1347 (2008).

\bibitem{Little1997_JLT}
B.~E. Little, S.~T. Chu, H.~A. Haus, \emph{et~al.}, \enquote{Microring resonator channel dropping filters,} {\protect\JournalTitle{J. Light. Technol.}} \textbf{15}, 998--1005 (1997).

\bibitem{Matsko2006_JSTQE}
A.~B. Matsko and V.~S. Ilchenko, \enquote{Optical resonators with whispering-gallery modes-part {I}: basics,} {\protect\JournalTitle{IEEE J. Sel. Top. Quantum Electron.}} \textbf{12}, 3--14 (2006).

\bibitem{Zhu_OE_2021}
S.~Zhu, X.~Ma, C.~Liu, \emph{et~al.}, \enquote{Controlled single-mode emission in quantum dot micro-lasers,} {\protect\JournalTitle{Opt. Express}} \textbf{29}, 13193--13203 (2021).

\bibitem{Jin_JOSAB_2016}
X.~Jin, Y.-D. Yang, J.-L. Xiao, and Y.-Z. Huang, \enquote{Mode control for microring resonators with inner-wall gratings,} {\protect\JournalTitle{J. Opt. Soc. Am. B}} \textbf{33}, 1906--1912 (2016).

\bibitem{Cai_Science_2012}
X.~Cai, J.~Wang, M.~J. Strain, \emph{et~al.}, \enquote{Integrated compact optical vortex beam emitters,} {\protect\JournalTitle{Science}} \textbf{338}, 363--366 (2012).

\bibitem{Chen_ACS_2024}
J.~Chen, A.~Nasir, A.~Abazi, \emph{et~al.}, \enquote{Coexistence of the radial-guided mode and {WGM} in azimuthal-grating-integrated microring lasers,} {\protect\JournalTitle{ACS Photonics}} \textbf{11}, 5110--5117 (2024).

\bibitem{Arbabi_OE_2015}
A.~Arbabi, S.~M. Kamali, E.~Arbabi, \emph{et~al.}, \enquote{Grating integrated single mode microring laser,} {\protect\JournalTitle{Opt. Express}} \textbf{23}, 5335--5347 (2015).

\bibitem{Feng_Science_2014}
L.~Feng, Z.~J. Wong, R.-M. Ma, \emph{et~al.}, \enquote{Single-mode laser by parity-time symmetry breaking,} {\protect\JournalTitle{Science}} \textbf{346}, 972--975 (2014).

\bibitem{Amano2006_APL}
T.~Amano, T.~Sugaya, and K.~Komori, \enquote{Characteristics of 1.3 $\mu$m quantum-dot lasers with high-density and high-uniformity quantum dots,} {\protect\JournalTitle{Appl. Phys. Lett.}} \textbf{89}, 171122 (2006).

\bibitem{Zenari2023_ACS}
M.~Zenari, M.~Buffolo, C.~De~Santi, \emph{et~al.}, \enquote{Addressing the optical degradation of 1.3 $\mu$m quantum dot lasers through subthreshold characterization,} {\protect\JournalTitle{ACS Photonics}} \textbf{10}, 4188--4195 (2023).

\bibitem{Ghosh2001_APL}
S.~Ghosh, P.~Bhattacharya, E.~Stoner, \emph{et~al.}, \enquote{Temperature-dependent measurement of {Auger} recombination in self-organized {In$_{0.4}$Ga$_{0.6}$As/GaAs} quantum dots,} {\protect\JournalTitle{Appl. Phys. Lett.}} \textbf{79}, 722--724 (2001).

\bibitem{Maximov2008_SST}
M.~V. Maximov, V.~M. Ustinov, A.~E. Zhukov, \emph{et~al.}, \enquote{A 1.33 $\mu$m {InAs/GaAs} quantum dot laser with a 46 cm$^{-1}$ modal gain,} {\protect\JournalTitle{Semicond. Sci. Technol.}} \textbf{23}, 105004 (2008).

\bibitem{Uvin2018_OE}
S.~Uvin, S.~Kumari, A.~D. Groote, \emph{et~al.}, \enquote{1.3 $\mu$m {InAs/GaAs} quantum dot {DFB} laser integrated on a {Si} waveguide circuit by means of adhesive die-to-wafer bonding,} {\protect\JournalTitle{Opt. Express}} \textbf{26}, 18302--18309 (2018).

\bibitem{Jung2018_ACS}
D.~Jung, Z.~Zhang, J.~Norman, \emph{et~al.}, \enquote{Highly reliable low-threshold {InAs} quantum dot lasers on on-axis (001) {Si} with 87{\%} injection efficiency,} {\protect\JournalTitle{ACS Photonics}} \textbf{5}, 1094--1100 (2018).

\bibitem{Telcordia_GR468}
O.~S. Gebizlioglu, \enquote{{GR-468-CORE}: Generic reliability assurance requirements for optoelectronic devices used in telecommunications equipment,} Generic Requirements GR-468-CORE, Issue 2, Telcordia Technologies, Piscataway, NJ (2004).

\bibitem{ILXLightwave_AN12}
{ILX Lightwave Corporation}, \enquote{The differences between threshold current calculation methods,} Application Note AN-12, Rev.\ 03, ILX Lightwave Corporation, Bozeman, MT (2009).

\bibitem{Nagarajan1992_JQE}
R.~Nagarajan, M.~Ishikawa, T.~Fukushima, \emph{et~al.}, \enquote{High speed quantum-well lasers and carrier transport effects,} {\protect\JournalTitle{IEEE J. Quantum Electron.}} \textbf{28}, 1990--2008 (1992).

\bibitem{Inoue2018_OE}
D.~Inoue, D.~Jung, J.~Norman, \emph{et~al.}, \enquote{Directly modulated 1.3 $\mu$m quantum dot lasers epitaxially grown on silicon,} {\protect\JournalTitle{Opt. Express}} \textbf{26}, 7022--7033 (2018).

\bibitem{yang2026_experimentaldesign}
X.~Yang, P.~Luong, Y.~Ramanujam, \emph{et~al.}, \enquote{Experimental design space exploration of ultra-low threshold hybrid {III-V/Si} quantum dot microring lasers,} {\protect\JournalTitle{arXiv preprint arXiv:2606.13371}}  (2026).

\end{thebibliography}

\end{document}